\documentclass[aps,prc,10pt,tightenlines,notitlepage,showpacs,nofootinbib,superscriptaddress,twocolumn,eqsecnum,raggedbottom,preprintnumbers,longbibliography]{revtex4-1}

\usepackage[export]{adjustbox}
\usepackage{amsfonts}
\usepackage{amsmath}
\usepackage{amssymb}
\usepackage{array}
\usepackage{bbold}
\usepackage{bm}
\usepackage{booktabs}
\usepackage[dvipsnames,usenames]{color}
\usepackage{enumitem}
\usepackage{float}
\usepackage{graphicx}
\usepackage[utf8]{inputenc}
\usepackage{lipsum}
\usepackage{multirow}
\usepackage{nicefrac}
\usepackage[normalem]{ulem}
\usepackage[dvipsnames]{xcolor}
\usepackage{hyperref}
\hypersetup{
    colorlinks=true,
    urlcolor=blue,
    linkcolor = blue,
    urlcolor  = blue,
    citecolor = blue,
    anchorcolor = blue,
    pdftitle={JBW-CC-electroproduction},
    pdfauthor={JBW},
    pdfsubject={JBW-CC-electroproduction}}

\newcommand{\overbar}[1]{\mkern 1.5mu\overline{\mkern-1.5mu#1\mkern-1.5mu}\mkern 1.5mu}

\begin{document}


\title{Coupled-channel analysis of pion- and eta-electroproduction with the J\"ulich-Bonn-Washington model}

\author{M.~Mai}
\affiliation{Institute for Nuclear Studies and Department of Physics, The George Washington University, Washington, DC 20052, USA}

\author{M.~D\"oring}
\affiliation{Institute for Nuclear Studies and Department of Physics, The George Washington University, Washington, DC 20052, USA}
\affiliation{Thomas Jefferson National Accelerator Facility, Newport News, VA 23606, USA}

\author{C.~Granados}
\affiliation{Institute for Nuclear Studies and Department of Physics, The George Washington University, Washington, DC 20052, USA}

\author{H.~Haberzettl}
\affiliation{Institute for Nuclear Studies and Department of Physics, The George Washington University, Washington, DC 20052, USA}

\author{J.~Hergenrather}
\affiliation{Institute for Nuclear Studies and Department of Physics, The George Washington University, Washington, DC 20052, USA}

\author{Ulf-G.~Mei{\ss}ner}
\affiliation{Helmholtz-Institut f\"ur Strahlen- und Kernphysik (Theorie) and Bethe Center for Theoretical Physics,  Universit\"at Bonn, 53115 Bonn, Germany}
\affiliation{Institute for Advanced Simulation and J\"ulich Center for Hadron Physics, Forschungszentrum J\"ulich,  52425 J\"ulich, Germany}
\affiliation{Tbilisi State University, 0186 Tbilisi, Georgia}

\author{D.~R\"onchen}
\affiliation{Institute for Advanced Simulation and J\"ulich Center for Hadron Physics, Forschungszentrum J\"ulich, 
52425 J\"ulich, Germany}

\author{I.~Strakovsky}
\affiliation{Institute for Nuclear Studies and Department of Physics, The George Washington University, Washington, DC 20052, USA}

\author{R.~Workman}
\affiliation{Institute for Nuclear Studies and Department of Physics, The George Washington University, Washington, DC 20052, USA}

\collaboration{J\"ulich-Bonn-Washington Collaboration}
\preprint{
JLAB-THY-21-3524}

\begin{abstract}
Pion and eta electroproduction data are  jointly analyzed for the first time, up to a center-of-mass energy of 1.6 GeV. The framework is a dynamical coupled-channel model, based on the recent J\"ulich-Bonn-Washington analysis of pion electroproduction data for the same energy range. Comparisons are made to a number of  single-channel eta electroproduction fits. By comparing multipoles of comparable fit quality, we find some of these amplitudes are well determined over the near-threshold region, while others will require fits over an extended energy range.
\end{abstract}



\maketitle   

\section{Introduction}
\label{sec:introduction}

Early progress in baryon spectroscopy was driven by the analysis of meson-nucleon scattering data, such as pion-nucleon scattering ($\pi N\to \pi N$, $\pi N\to \pi \pi N$); see, e.g., Refs~\cite{Hohler:1984ux, Cutkosky:1979zv, Arndt:2006bf, Arndt:2009nv, Shrestha:2012ep}. A viable alternative to this, specifically advantageous for detecting unstable intermediate states with small branching ratios to the $\pi N$ channel, are photon-induced reactions~\cite{Ireland:2019uwn}. Large data bases have been accumulated for these reactions as part of the extensive experimental programs at Jefferson Laboratory, MAMI, ELSA and other facilities~\cite{Aznauryan:2009mx, Achenbach:2017pse, Carman:2020qmb, Beck:2017wkb, Shiu:2017wiw, A2:2018doh,CLAS:2018gxz,Ohnishi:2019cif, Alef:2019imq,CLAS:2021udy,CLAS:2021osv,CLAS:2019cpp}. 

On the theory side, many approaches have been developed to describe these reactions. Specifically for pion electroproduction, chiral perturbation theory (ChPT) has been successfully applied in the analysis of the threshold region~\cite{Bernard:1992ms, Bernard:1992rf, Bernard:1993bq, Bernard:1996bi,Hilt:2013fda}. In building on ChPT, chiral unitary models (see, e.g., the recent review~\cite{Mai:2020ltx}) have also become quite successful in accessing the resonance region. Such models provide the hadronic structure of many gauge-invariant chiral unitary formalisms~\cite{Ruic:2011wf, Mai:2012wy, Bruns:2020lyb, Doring:2009qr, Doring:2009uc, Borasoy:2007ku, Meissner:1999vr}. For an in-depth discussion of the manifest implementation of gauge invariance see Ref.~\cite{Haberzettl:2021wcz}. For even larger kinematical ranges, and large data bases, many phenomenological models have been developed. Major classes of those are: (1) isobar models~\cite{Chiang:2001as, Aznauryan:2002gd,Drechsel:2007if, Tiator:2018heh} with unitarity constraints at lower energies; (2) $K$-matrix-based formalisms with built-in cuts associated with opening inelastic channels, and dispersion-relation constraints~\cite{Aznauryan:2002gd, Hanstein:1997tp, Workman:2012jf}.  Multi-channel analyses have analyzed data and, in some cases, amplitudes from hadronic scattering data together with the photon-induced channels~\cite{Briscoe:2015qia} by the Gie{\ss}en~\cite{Shklyar:2006xw}, Bonn-Gatchina~\cite{Anisovich:2011fc}, Kent State~\cite{Shrestha:2012ep}, ANL-Osaka~\cite{Kamano:2013iva}, J\"ulich-Bonn (J\"uBo)~\cite{Ronchen:2015vfa} and JPAC~\cite{Nys:2016vjz} groups. For more details, see the introduction of our recent paper~\cite{Mai:2021vsw}.

Dynamical coupled-channel approaches~\cite{Ronchen:2018ury, Kamano:2016bgm, Ronchen:2015vfa, Kamano:2013iva, Ronchen:2014cna, Kamano:2010ud, Tiator:2010rp} (DCC) have led to the discovery and confirmation of many new states~\cite{Zyla:2020zbs} by extracting universal resonance parameters in terms of pole positions and residues of the transition amplitude in the complex-energy plane. In many analyses of pion- and photon-induced reactions, mass scans and $\chi^2$ arguments are used to identify new states~\cite{CBELSATAPS:2011nwh, Anisovich:2015gia} but recently  model selection has also been explored~\cite{Landay:2018wgf, Landay:2016cjw}. 

Furthermore, by extracting~\cite{ Drechsel:1998hk, Arndt:2001si, Chiang:2001as, Tiator:2003xr, Tiator:2003uu, Aznauryan:2004jd, Aznauryan:2005tp, Corthals:2007kc, Drechsel:2007if, Aznauryan:2009mx, Aznauryan:2011qj, Tiator:2011pw, Aznauryan:2012ba,  Vrancx:2014pwa, Maxwell:2014txa, Mokeev:2015lda, Isupov:2017lnd, Burkert:2019opk, Blin:2019fre, Mart:2002gn, Mokeev:2020hhu} the $Q^2$ dependence of resonance couplings, a link between perturbative QCD and the region where quark confinement sets in can be established that serve as point of comparison for many quark models~\cite{Isgur:1978xj,Capstick:1986bm, Capstick:1992uc, Capstick:1993kb,Ronniger:2011td, Ramalho:2011ae,Jayalath:2011uc,  Aznauryan:2012ec, Golli:2013uha, Obukhovsky:2019xrs, Ramalho:2019koj, Ramalho:2020nwk} and Dyson-Schwinger equations~\cite{Roberts:1994dr,Roberts:2007ji, Eichmann:2009qa, Wilson:2011aa, Chen:2012qr, Eichmann:2012mp, Xu:2015kta, Segovia:2015hra, Eichmann:2016yit, Eichmann:2016hgl, Burkert:2017djo, Chen:2018nsg, Qin:2018dqp,Qin:2019hgk, Chen:2019fzn, Lu:2019bjs}.

However, so far, no unified coupled-channel analysis of photo- and electroproduction experiments exists that simultaneously describes the $\pi N$, $\eta N$ and $K\Lambda$ final states. The present study provides a first step in this direction in the form of a coupled-channel analysis of pion and eta electroproduction data, extending our recent analysis of pion electroproduction data~\cite{Mai:2021vsw}. It is based on the J\"uBo approach~\cite{Ronchen:2015vfa} which fits an extensive scattering and photoproduction data base in the resonance region.

This study is organized as follows. Section II outlines formal aspects of the J\"ulich-Bonn-Washington (JBW) approach to pseudo-scalar meson electroproduction. These include the generalization to  electroproduction of different final states and the influence of kinematic limits ($Q^2=0$, thresholds for channel openings, and pseudo-thresholds associated with Siegert's theorem~\cite{Siegert:1937yt, Tiator:2016kbr}). Furthermore, we define the parametrization of the  $Q^2$ dependence from the photon point at $Q^2=0$, where the underlying J\"uBo model describes photon- and meson-induced reactions, to electroproduction. 

Section~\ref{sec:single_channel} reviews the results of previous single-channel fits to eta electroproduction data.
Section~\ref{sec:expdata} describes the data used in our fits, strategies to find $\chi^2$ minima, and a modified $\chi^2$ which more evenly weights the contributions of observables with different abundances. Section~\ref{sec:results} compares our fits to data. Eta electroproduction multipoles are also compared, with etaMAID results displayed for reference. 

Section~\ref{sec:conclusion} compares the single- and coupled-channel results, with qualitative features discussed based on data quality/consistency. Finally, prospects for expanded analyses are considered.

\section{Formalism}
\label{sec:formalism}

\begin{figure}[t]
\centering
\includegraphics[width=\linewidth,trim=0 7cm 12.6cm 7cm,clip]{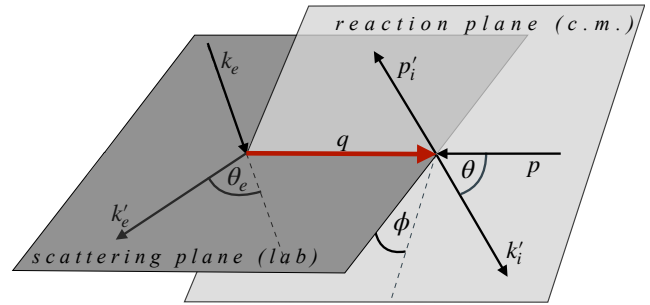}
\caption{
\label{fig:kinematics}
Kinematics of an electroproduction experiment with the final meson-baryon state $i$. The scattering plane is defined by the respective in/outgoing electron momenta $k_e/k_e'$ with the electron scattering angle $\theta_e$. The reaction plane is spanned by the virtual photon and the outgoing meson, scattered by an angle $\theta$.
}
\end{figure}

The multichannel meson electroproduction process\footnote{Some parts of the present section overlap with the discussion in Ref.~\cite{Mai:2021vsw}, but here we generalize the framework to any final meson-baryon state.} in question reads 
\begin{align}
\gamma^*(\bm{q})+p(\bm{p})
\to
M(\bm{k}'_i)+B(\bm{p}'_i)\,,
\end{align}
where bold symbols denote three-momenta  throughout the manuscript. The meson and baryon in the final state, with the index $i$, are denoted by $M$ and $B$, respectively. As shown in Fig.~\ref{fig:kinematics}, the process occurs in two steps, with a virtual photon $\gamma^*(\bm{q})$ being produced via $e_{\rm in}(\bm{k}_e)\to e_{\rm out}(\bm{k}'_e)+\gamma^*(\bm{q})$, which then scatters off of the proton to a final meson-baryon state. The momentum transfer $Q^2=-\omega^2+\bm{q}^2$, where $\omega$ is the photon energy, is non-negative for spacelike processes, and acts as an independent kinematical variable in addition to the total energy in the center-of-mass (cms) frame, $W$. In this frame, the magnitude of the three momentum of the photon ($q=|\bm{q}|$) and produced meson ($k'=|\bm{k}'|$) read
\begin{align}
q=\frac{\sqrt{\lambda(W^2,m_p^2,-Q^2)}}{2W}\,,~
k_i'=\frac{\sqrt{\lambda(W^2,m_i^2,M_i^2)}}{2W}\,,
\label{eq:qk}
\end{align}
where $\lambda(x,y,z)=x^2+y^2+z^2-2xy-2yz-2zx$ denotes the usual K\"all\'en triangle function. The meson and baryon masses are denoted throughout this manuscript by $M_i$ and $m_i$, respectively.
With two incoming and three outgoing states there are $(3+2)\times3-10=5$ independent kinematic variables. The canonical choice for the remaining three (in addition to $W$ and $Q^2$) variables is illustrated in Fig.~\ref{fig:kinematics}. The quantity $\epsilon=1+2({q_L^2}/{Q^2})\tan^2{\theta_e}/{2}$ contains the electron scattering angle $\theta_e$ and $q_L$ denotes the photon three-momentum in the laboratory frame. The angle of the reaction plane to the scattering plane is given by $\phi$ and $\theta$ is the c.m.\ meson scattering angle in the latter plane. The experimental data discussed in the Sec.~\ref{sec:expdata} are represented with respect to these five variables $\mathcal{O}(Q^2,W,\phi,\theta,\epsilon)$.

As discussed in the previous paper~\cite{Mai:2021vsw} based on the seminal papers~\cite{Chew:1957tf,Dennery:1961zz,Berends:1967vi,CiofiDegliAtti:1981wk}, the process of a  photon-induced production of a meson off a nucleon is encoded in the transition amplitude. In the one-photon approximation, and considering the continuity equation for the current, the latter can be expressed in terms of three independent multipoles for a fixed quantum number $\ell_\pm$ of the final meson-baryon state. We chose those to be electric, magnetic and longitudinal multipoles $E_{\ell\pm}(W,Q^2)$, $M_{\ell\pm}(W,Q^2)$ and $L_{\ell\pm}(W,Q^2)$ with the latter related to the often-used Coulomb multipole as $\omega C_{\ell\pm}(W,Q^2)=q L_{\ell\pm}(W,Q^2)$. Each of the introduced multipoles carry a discrete index corresponding to the total angular momentum $J=\ell\pm 1/2$ and final-state index $\mu$, e.g., $E^{\eta p}_{0+}$.

We construct the electroproduction multipoles on the basis of the dynamical coupled-channel J\"ulich-Bonn (J\"uBo) approach~\cite{Ronchen:2012eg,Ronchen:2014cna} that provides the boundary condition at $Q^2=0$, incorporating the experimental information from real-photon and pion-induced reactions. In this approach, two-body unitarity and analyticity are respected and the baryon resonance spectrum is determined in terms of poles in the complex energy plane on the second Riemann sheet~\cite{Doring:2009yv, Doring:2009bi}.

Extending the ansatz of the J\"uBo approach, we begin by introducting a generic function ($\bar{\cal{M}}$) for each electromagnetic multipole  (${\cal M}_{\mu\gamma^*}\in \{E_\mu,L_\mu,M_\mu\}$) as
\begin{align}
\label{eq:m_electro}
    \overbar{\cal M}_{\mu\gamma^*}&(k,W,Q^2)=V_{\mu\gamma^*}(k,W,Q^2)\\
    +&\sum_\kappa\int\limits_0^\infty dp\, p^2\, T_{\mu\kappa}(k,p,W)G_\kappa(p,W)V_{\kappa\gamma^*}(p,W,Q^2)\,,\nonumber
\end{align}
where the summation extends over intermediate meson-baryon channels $\kappa\in\{\pi N,\eta N, K\Lambda, K\Sigma, \pi\Delta, \rho N\}$. Note that the $\sigma N$ channel is not part of this list. The $\sigma N$ channel is part of the final-state interaction, but neither the hadronic resonance vertex functions nor the photon is directly coupled to it; once two-pion photon or electroproduction data are analyzed, such couplings will become relevant and will be included.

The electroproduction kernel $V_{\mu\gamma^*}$ in Eq.~\eqref{eq:m_electro} is parametrized as
\begin{align}
\label{eq:v_electro}
    V_{\mu\gamma^*}(p,W,Q^2)=&
    \alpha^{NP}_{\mu\gamma^*}(p,W,Q^2)\\\nonumber
    &+\sum_{i=1}^{i_\text{max}}\frac{\gamma^a_{\mu;i}(p)\gamma^c_{\gamma^*;i}(W,Q^2)}{ W-m^b_i}\,,
\end{align}
introducing the $Q^2$-dependence via a separable ansatz,
\begin{align}
    \alpha^{NP}_{\mu\gamma^*}(p,W,Q^2) &={\tilde F}_\mu(Q^2)\alpha^{NP}_{\mu\gamma}(p,W)\,,\nonumber\\
    \gamma^c_{\gamma^*;i}(W,Q^2)       &={\tilde F}_i(Q^2)\gamma^c_{\gamma;i}(W)\,.
\label{eq:ff_electro_2}
\end{align}
The $Q^2$-independent pieces on the right hand side of both equations represent the input from the J\"uBo2017 solution~\cite{Ronchen:2018ury}. Specifically, $\gamma^c_{\gamma;i}$ describes the interaction of the photon with the resonance state $i$ with bare mass $m_i^b$ and $\alpha^{NP}_{\mu\gamma}$ accounts for the coupling of the photon to the so-called background or non-pole part of the amplitude. Both quantities are parametrized by energy-dependent polynomials, see Ref.~\cite{Ronchen:2014cna}.

The $Q^2$ dependence is encoded entirely in the channel-dependent form-factor ${\tilde F}_\mu(Q^2)$ and another channel-independent form-factor ${\tilde F}_i(Q^2)$ that depends on the resonance index $i$. We emphasize that this structure is inherited from the J\"uBo photoproduction ansatz, which separates the photon-induced vertex ($\gamma^c$) from the decay vertex of a resonance to the final meson-baryon pair ($\gamma_\mu^a$). Both ${\tilde F}_\mu(Q^2)$ and ${\tilde F}_i(Q^2)$ are chosen as
\begin{align}
    {\tilde F}_\mu(Q^2)&={\tilde F}_D(Q^2)\,e^{-\beta_\mu^0 Q^2/ m^2}\,P^N(Q^2/m^2,\vec{\beta}_\mu)\,,\nonumber\\
    {\tilde F}_i(Q^2)&={\tilde F}_D(Q^2)\,e^{-\delta_i^0 Q^2/ m^2}\,P^N(Q^2/m^2,\vec{\delta}_i)\,,
    \label{eq:formfactor-Ftilde}
\end{align}
where $P^N(x,\vec y)= 1 + xy_1 +...+x^Ny_N$ is a general polynomial with free parameters to be fitted together with $\delta_i^0$ and $\beta^0_\mu$ to experimental electroproduction data. The parameter-free form factor ${\tilde F}_D(Q^2)$ encodes the empirical dipole behavior, usually implemented in such problems, as well as a Woods-Saxon form factor which ensures suppression at large $Q^2$. It reads
\begin{align}
    \label{eq:formfactor-FD}
    {\tilde F}_D(Q^2)=\frac{1}{(1+Q^2/b^2)^{2}}\,
    \frac{1+e^{-Q_r^2/Q_w^2}}{1+e^{(Q^2-Q_r^2)/Q_w^2}}
\end{align}
with $b^2=0.71$~GeV${}^2$, $Q_w^2=0.5~{\rm GeV}^2$ and  $Q_r^2=4.0~{\rm GeV}^2$, see Ref.~\cite{Mai:2021vsw} for more details

As stated above, this procedure relies heavily on the input from the photoproduction, i.e., the functions $\alpha^{NP}_{\mu \gamma}(p,W)$ and $\gamma^c_{\gamma;i}(W)$. Obviously, such an input cannot exist for the longitudinal multipoles as their contribution vanishes exactly at the photon-point. In this case we employ a strategy similar to that of Ref.~\cite{Mai:2012wy}:

1) We recall that at the \emph{pseudo-threshold} ($q=0$) the electric and longitudinal multipoles relate according to the Siegert's condition as
\begin{equation}
    \frac{E_{\ell+}}{L_{\ell+}}\Big|_{q=0}=1\,,\qquad
    \frac{E_{\ell-}}{L_{\ell-}}\Big|_{q=0}=\frac{\ell}{1-\ell}\,.
\label{eq:Siegerts_condition}
\end{equation}
For more details, see Sec.~2.2-2.3 of Ref.~\cite{Mai:2012wy}, or the original derivations in Refs.~\cite{CiofiDegliAtti:1981wk,Tiator:2016kbr}. Therefore, we apply at the nearest  pseudo-threshold point, $Q^2_{\rm PT}=-(W-m)^2$,
\begin{align}
\alpha^{NP,L_{\ell\pm}}_{\mu\gamma^*}(p,W,Q^2)&=
\frac{\omega}{\omega_{\rm PT}}
\frac{{\tilde F}_D(Q^2)}{{\tilde F}_D(Q^2_{\rm PT})}
\label{pt_cond3}\\
&\times D_\mu^{\ell\pm}(W,Q^2)
\alpha^{NP,E_{\ell\pm}}_{\mu\gamma^*}(p,W,Q^2_{\rm PT})\,,\nonumber
\end{align}
and
\begin{align}
\gamma^{c,L_{\ell\pm}}_{\gamma^*;i}(W,Q^2)=
\frac{\omega}{\omega_{\rm PT}}&
\frac{{\tilde F}_D(Q^2)}{{\tilde F}_D(Q^2_{\rm PT})}
\label{pt_cond4}\\
&\times\tilde D_i^{\ell\pm}(W,Q^2)
\gamma^{c,E_{\ell\pm}}_{\gamma^*;i}(W,Q^2_{\rm PT})\,.\nonumber
\end{align}
The photon energy is $\omega_{\rm PT}=(W^2-m^2-Q_{\rm PT}^2)/(2W)$. The new functions $D^{\ell\pm}(Q^2)$ ensure Siegert's condition and consistent falloff behavior in $Q^2$ as
\begin{align}
D_\mu^{\ell+}(W,Q^2)&=
    e^{-\beta_\mu^0 q/q_\gamma}\,P^{N}(q/q_\gamma,\vec\beta_\mu)\,,\label{eq:D-formfactor}\\
\tilde D_i^{\ell+}(W,Q^2)&=
    e^{-\delta_i^0 q/q_\gamma}\,P^{N}(q/q_\gamma,\vec\delta_i)\,,\nonumber\\
D_\mu^{\ell-}(W,Q^2)&=
    -\frac{\ell-1}{\ell}e^{-\beta_i^0 q/q_\gamma}\,P^{N}(q/q_\gamma,\vec\beta_\mu)\,,\nonumber\\
\tilde D_i^{\ell-}(W,Q^2)&=
    -\frac{\ell-1}{\ell}e^{-\delta_i^0 q/q_\gamma}\,P^{N}(q/q_\gamma,\vec\delta_i)\,,\nonumber
\end{align}
respectively to the pole and non-pole part for $q_\gamma=q(Q^2=0)$.

2) In two specific cases ($(\ell\pm,I)=(1-,1/2)$ and $(\ell\pm,I)=(1-,3/2)$) the electric multipole vanishes due to selection rules, rendering the implementation of Siegert's theorem nonsensical. In these cases, we obtain the longitudinal multipole from the magnetic one using a new real-valued normalization constants $\zeta^{NP}$ to be determined from the fit,
\begin{align}
\label{eq:zetanp}
\alpha^{NP,L_{\ell\pm}}_{\mu\gamma^*}(p,W,Q^2)&=\zeta^{NP}_\mu\frac{\omega}{\omega_{\rm PT}}{\tilde F}^\mu(Q^2)\\
&\qquad\times \alpha^{NP,M_{\ell\pm}}_{\mu\gamma^*}(p,W)\,,\nonumber\\
\gamma^{c,L_{\ell\pm}}_{\gamma^*;i}(W,Q^2)&=\zeta_i\frac{\omega}{\omega_{\rm PT}}{\tilde F}^\mu(Q^2)\gamma^{c,M_{\ell\pm}}_{\gamma;i}(W)\,.\nonumber
\end{align}

Before writing down the final relation between the generic multipole functions ($\bar E_{\ell\pm}$, $\bar M_{\ell\pm}$, $\bar L_{\ell\pm}$) and corresponding multipoles, we note that the latter obey a certain behavior at the pseudo- ($q=0$) and production threshold ($k=0$)
\begin{align}
&\ell\geq0: \quad\lim_{k\to 0}E_{\ell+}=k^\ell~, &&\lim_{q\to 0}E_{\ell+}=q^\ell~,\nonumber\\
&\ell\geq0: \quad\lim_{k\to 0}L_{\ell+}=k^\ell~, &&\lim_{q\to 0}L_{\ell+}=q^\ell~,\nonumber\\
&\phantom{\ell=1:}    \quad\lim_{k\to 0}L_{1-}=k~, &&\lim_{q\to 0}L_{1-}=q~,\nonumber\\
&\ell\geq1: \quad\lim_{k\to 0}M_{\ell\pm}=k^\ell~, &&\lim_{q\to 0}M_{\ell\pm}=q^\ell~,\nonumber\\
&\ell\geq2: \quad\lim_{k\to 0}E_{\ell-}=k^\ell~, &&\lim_{q\to 0}E_{\ell-}=q^{\ell-2}~,\nonumber\\
&\ell\geq2: \quad\lim_{k\to 0}L_{\ell-}=k^\ell~, &&\lim_{q\to 0}L_{\ell-}=q^{\ell-2}~.
\label{eq:pseudo-thre-conditions}
\end{align}
We incorporate these conditions using 
\begin{align}
    {\cal M}_{\mu\gamma^*}(k,W,Q^2)=
    R_{\ell'}(\lambda, q/q_\gamma)\overbar{\cal M}_{\mu\gamma^*}(k,W,Q^2)
    \label{ampl_2}
\end{align}
for each multipole type and total angular momentum individually. Here,
\begin{align}
    R_{\ell'}(\lambda,r)&=\frac{B_{\ell'}(\lambda r)}{B_{\ell'}(\lambda)}
    \\
    &\text{with}\quad\ell'=
\left\{
\begin{matrix}
\ell,~~\text{for}~E_{\ell+},L_{\ell\pm},M_{\ell\pm}~,~~~\\
\ell-2,~~\text{for}~E_{\ell-},L_{\ell-}\text{  and  }\ell\geq 2\,,
\end{matrix}
\right. \nonumber
\end{align}
using Blatt-Weisskopf barrier-penetration factors~\cite{Blatt:1952,Manley:1984jz},
\begin{align}\label{BW_fctr}
&B_0(r)=1\,,\\
&B_1(r)=r/\sqrt{1+r^2}\,,\nonumber\\
&B_2(r)=r^2/\sqrt{9+3r^2+r^4}\,,\nonumber\\
&B_3(r)=r^3/\sqrt{225+45r^2+6r^4+r^6}\,,\nonumber\\
&B_4(r)=r^4/\sqrt{11025+1575r^2+135r^4+10r^6+r^8}\, .\nonumber
\end{align}
New free parameters $\lambda$ need to be determined from a fit to experimental data. For simplicity and to keep the number of parameters low, the $\lambda$s are chosen as channel independent.

In summary, for every partial wave, the multipoles $E_{\mu\gamma^*}$, $M_{\mu\gamma^*}$ and $L_{\mu\gamma^*}$ are fully determined up to: (1) $(1+N)$ channel-dependent fit parameters $\beta^0_\mu,...,\beta^N_\mu$ for the non-pole part; (2) $(1+N)$ channel-independent parameters $\delta^0_i,...,\delta^N_i$ for each of the $i_\text{max}$ resonances; (3) one channel-independent threshold behavior regulating parameter $\lambda$; (4) channel-(in)dependent normalization factors $\zeta^{NP}_\mu(\zeta_i)$. Finally, any observable can be constructed from the described multipoles using a standard procedure involving CGLN and helicity amplitudes~\cite{Chew:1957tf}. For explicit formulas  we refer the reader to the previous publication~\cite{Mai:2021vsw}.


\section{Previous single-channel fits}
\label{sec:single_channel}

\begin{figure}[t]
    \includegraphics[width=\linewidth,trim=0cm 0 0.cm 0,clip]{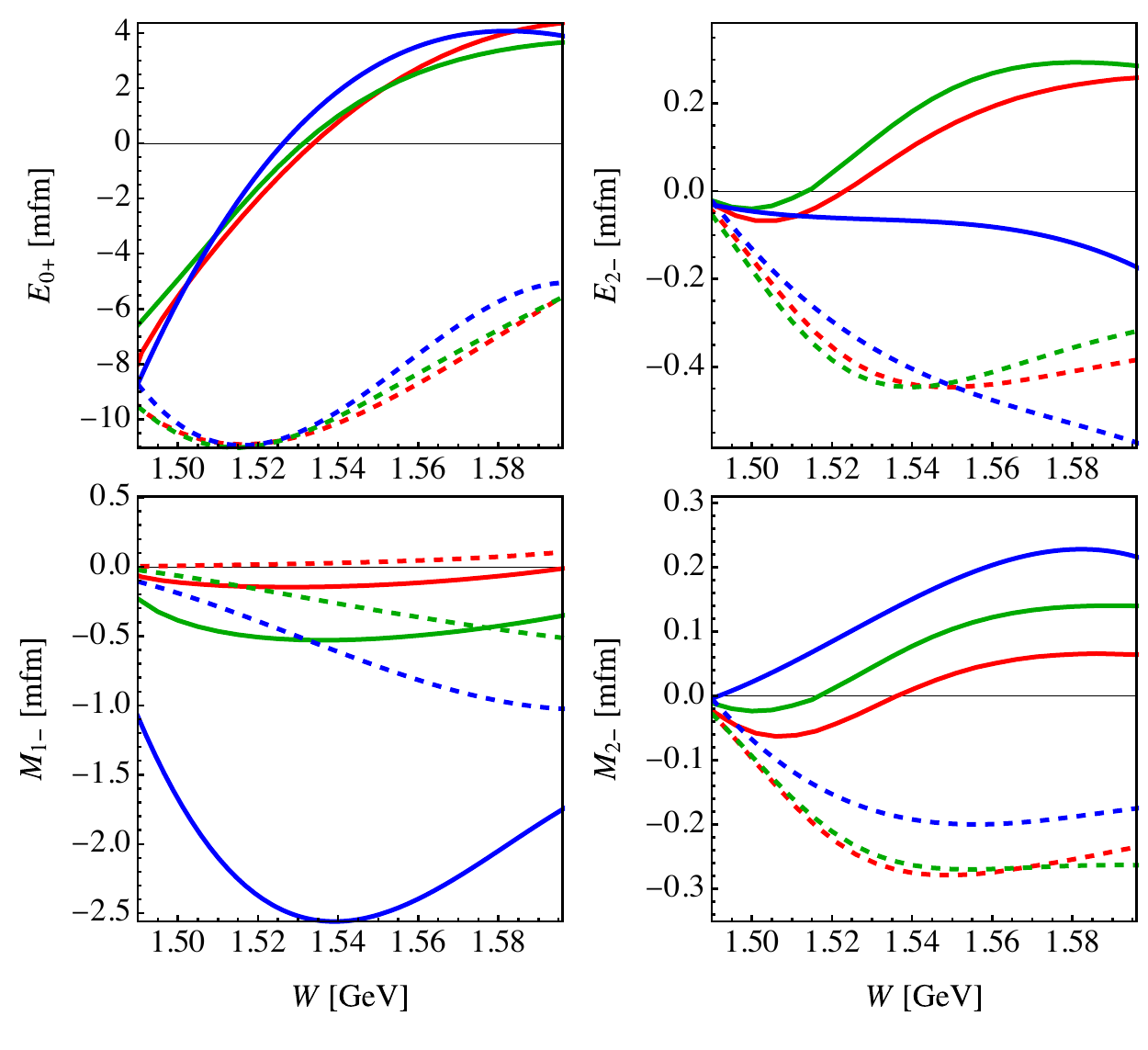}
    \caption{Selected eta photoproduction multipoles in the isospin basis from J\"uBo~\cite{Ronchen:2018ury} (blue), etaMAID~\cite{Chiang:2001as} (red) and Bonn-Gatchina~\cite{Anisovich:2012ct} (green) approaches. Real and imaginary parts are depicted by full and dashed lines, respectively. A phase factor $(-1)$ is applied to the etaMAID solution.
    \label{fig:eta-mult-Q0}}
\end{figure}

Before discussing our findings for the coupled-channel case, we review what has been learned from fits to the eta electroproduction data alone. The etaMAID model~\cite{Chiang:2001as} is similar to the MAID2007 analysis~\cite{Drechsel:2007if} of pion production, with the fit including both eta photo- and electroproduction data. It differs from MAID2007 by a phase factor which was adjusted to the corresponding pion-nucleon phase. For eta photo- and electroproduction this was not found to be feasible, due to the quality and range of available $\eta N$ production data. As a result, there exists an overall phase ambiguity, in comparisons of different eta photoproduction fits, which cannot be determined experimentally. Comparing the eta photoproduction fits of etaMAID and the J\"ulich-Bonn approach, overall qualitative agreement is improved by applying a simple overall sign. This is illustrated in Fig.~\ref{fig:eta-mult-Q0}. 
The overall phase, applied to etaMAID, yeilds a quantitative agreement between the J\"uBo, etaMAID, and Bonn-Gatchina determinations of the $E_{0+}$ multipole. The $M_{2-}$ multipole shows qualitative agreement, while the $M_{1-}$ and $E_{2-}$ multipoles show differences in sign and scale that hinder a comparison of electroproduction results extrapolated to $Q^2=0$.

\begin{figure*}[t]
\centering
\includegraphics[height=6.5cm,trim= 0 0cm 0 0,clip]{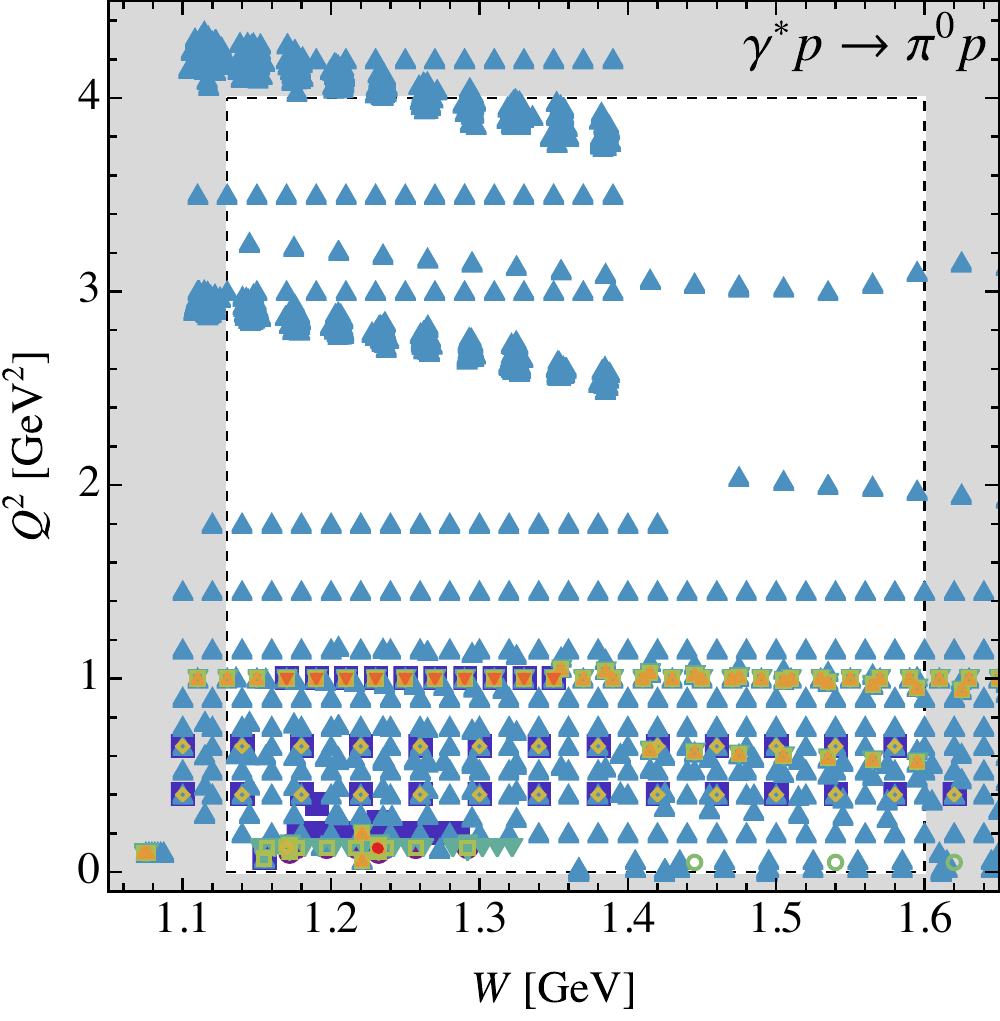}
\includegraphics[height=6.5cm,trim= 1.1cm 0cm 0 0,clip]{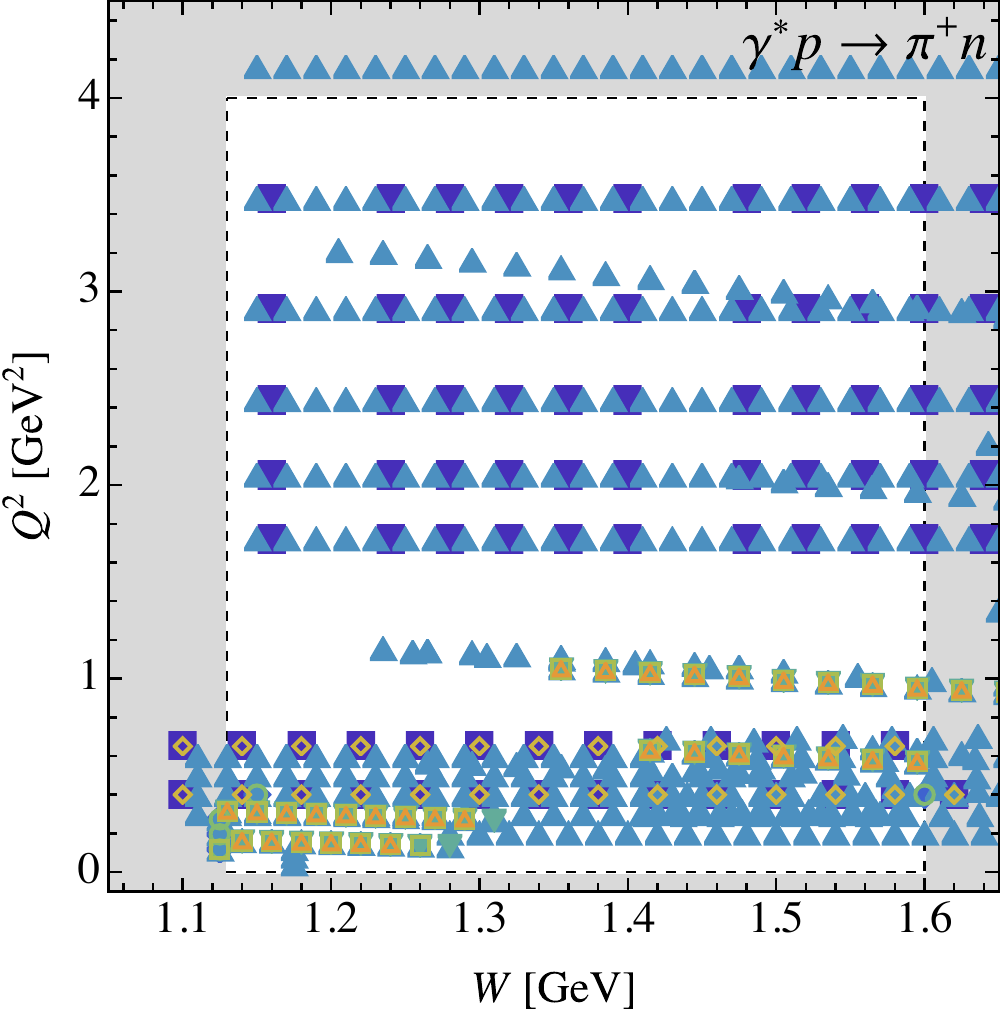}
\includegraphics[height=6.5cm,trim= 1.1cm 0cm 0 0,clip]{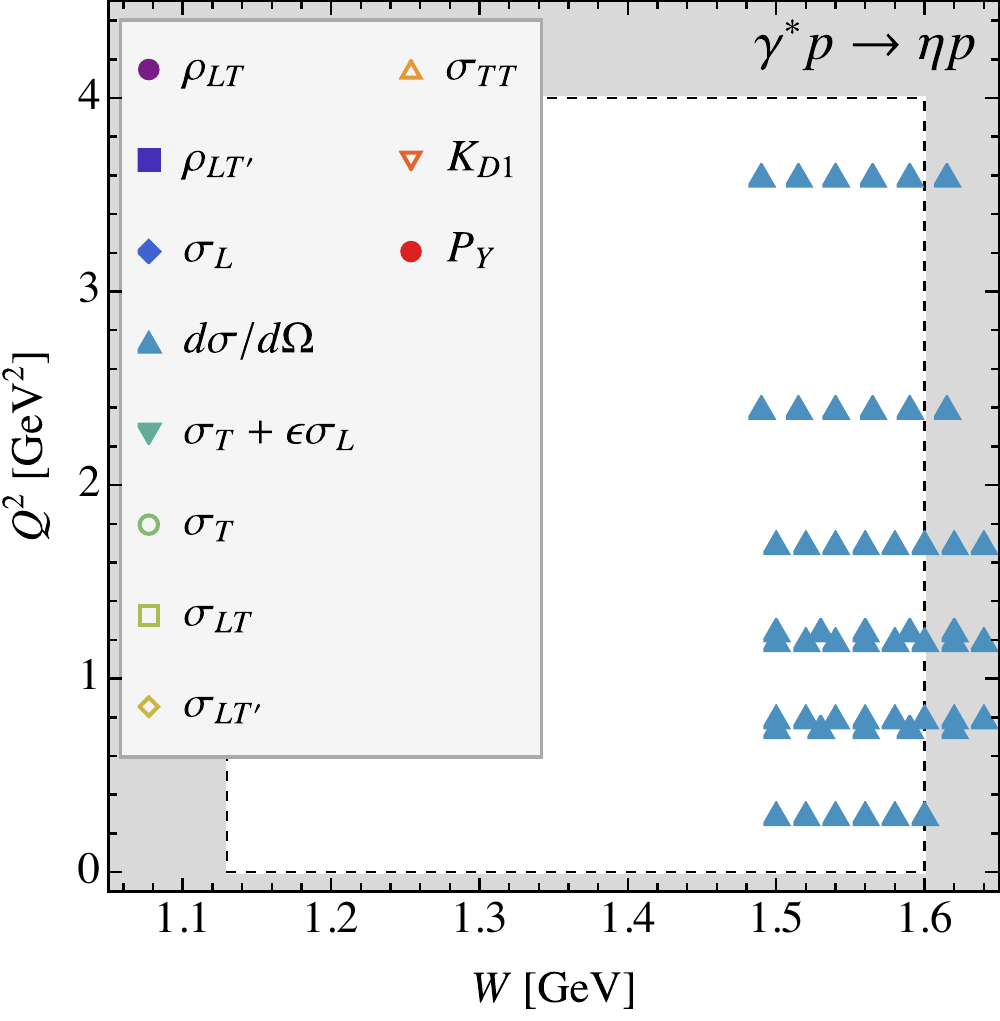}
\caption{
\label{fig:data}
Experimental data~\cite{Mertz:1999hp, Elsner:2005cz, Joo:2003uc, Sparveris:2002fh, Kelly:2005jj, Bartsch:2001ea, Bensafa:2006wr, Laveissiere:2003jf, Ungaro:2006df, Gayler:1971zz, May:1971zza, Hill:1977sy,  Joo:2001tw, Frolov:1998pw, Siddle:1971ug, Haidan:1979yqa, Kalleicher:1997qf, Baetzner:1974xy, Latham:1979wea, Latham:1980my, Stave:2006jha, Sparveris:2006uk, Alder:1975xt, Afanasev:1975qa,  Shuttleworth:1972nw, Blume:1982uh, Rosenberg:1979zm, Gerhardt:1979zz, Kunz:2003we, Stave:2006ea, Sparveris:2004jn, Warren:1999pq, Pospischil:2000ad, Joo:2004mi, PARK-pc-08-2007, Gaskell:2001fn,Egiyan:2006ks, Breuker:1977vy, Bardin:1975oea, Bardin:1977zu, DavenportMartyn1980Eona, Vapenikova:1988fd, Alder:1975na, Evangelides:1973gg, Breuker:1982nw, Breuker:1982um, Alder:1975na,Armstrong:1998wg, Denizli:2007tq, Thompson:2000by} used in the present analysis (aggregated values of $\theta,\,\phi,\,\epsilon$). The white dashed rectangles represent the considered fitting window $(0<Q^2/{\rm GeV}^2<4, 1.13<W/{\rm GeV}<1.6)$.
}
\end{figure*}

The eta electroproduction fit of etaMAID requires a determination of the $Q^2$ dependence, which is chosen to be simpler than what was used for pion electroproduction (MAID2007). The dominant S-wave multipole near threshold, in principle, includes both the $N(1535)$ and $N(1650)$ resonance contributions. These have been combined using a single-quark transition model~\cite{Burkert:1992yk}. For the $N(1535)$ multipole, the $Q^2$ dependence is assumed to be proportional to a dipole form factor multiplied by a ratio of linear functions of $Q^2$. Other resonance multipoles have $Q^2$ dependence approximated by a simple dipole factor
multiplied by a ratio of kinematic factors.

Included in etaMAID are the above mentioned $N(1535)$ and $N(1650)$, together with the $N(1520)$, $N(1675)$, $N(1700)$, $N(1710)$, and $N(1720)$. Of these, the $N(1650)$, $N(1675)$, $N(1710)$, and $N(1720)$ were found to have $\eta N$ branching ratios at the 3 - 26\% level; the $N(1520)$, $N(1680)$, and $N(1700)$ contributed with branching ratios less than 1\%. 
The $N(1535)$ had a 50\% branching to the $\eta N$ channel; there is universal agreement on this branching to the 10\% level. 
Included in the fit were data available as of 2001, the cross section measurements of Ref.~\cite{Armstrong:1998wg} and Ref.~\cite{Thompson:2000by}. 

The data of Ref.~\cite{Armstrong:1998wg} were taken for $Q^2$ values of 2.4 and 3.6 GeV$^2$, and for c.m.\ energies between approximately 1.5 and 1.6 GeV. Cross sections show, within uncertainties, a flat angular distribution, independent of the angle $\phi$. Based on this and a relativistic quark model
expectation~\cite{Ravndal:1971cuf} that longitudinal contributions should be small, a Breit-Wigner-plus-background fit was done to extract the $N(1535)$ contribution. The authors of Ref.~\cite{Ravndal:1971cuf} concluded that the background was consistent with zero and terms beyond an S-wave approximation amounted to less than 7\%. The size of longitudinal contributions had also been explored experimentally~\cite{Brasse:1977as, Breuker:1978qr} by varying $\epsilon$ in order to extract the ratio of longitudinal and transverse cross sections. This ratio for $Q^2$ values between 0.4 and 1.0 GeV$^2$, and c.m. energies corresponding to the $N(1535)$, was found to be about 20\% with 100\% uncertainties. Cross sections up to a c.m. energy of 1.9 GeV, for $Q^2$ between 0.15 to 1.5 GeV$^2$, were measured in Ref~\cite{Thompson:2000by}. Here, the cross section was fitted to an expansion in multipoles up to $J = 3/2$. Unlike Ref.~\cite{Armstrong:1998wg}, evidence for significant interference between S- and P-wave contributions was found with cross sections displaying $\theta$ dependence also at a c.m.\ energy corresponding to the $N(1535)$.

Data from Ref.~\cite{Denizli:2007tq} have the benefit of a wide kinematic range, with c.m. energies between
1.5 and 2.3 GeV, and $Q^2$ values from 0.13 to 3.3 GeV$^2$. Cross sections were expanded in terms of Legendre polynomials. Here too, the angular behavior was concluded to be mainly due to interference between S- and P-wave contribution, though they could not distinguish between the $N(1710)$ and $N(1720)$ as a source. Fits to this data set tended to show more $\phi$ dependence in the cross section near the $N(1535)$ than was given in the etaMAID result -- though these data were not included in the etaMAID solution.

Data from Ref.~\cite{Dalton:2008aa} covers the c.m. energy range from threshold to 1.8 GeV for $Q^2$ values of 5.7 and 7.0 GeV$^2$. Angular behavior is again attributed to S- and P-wave interference. There is evidence that this interference may change sign in going from lower to higher values of $Q^2$. EtaMAID appears to give a reasonable qualitative description of the data even at these high $Q^2$ values, even though these data were not included in the fit.

\section{Data and fits}
\label{sec:expdata}

By design, the introduced framework is capable of addressing the electroproduction multipoles and observables simultaneously in all considered channels $\{\pi N,\eta N, K\Lambda, K\Sigma, \pi\Delta, \rho N\}$. Experimentally, the most extensively explored final-state channels are $\pi N$, $\eta N$, and $K\Lambda$. Therefore, and also extending upon the already available single channel JBW analysis of Ref.~\cite{Mai:2021vsw}, we first restrict the data base to pion and eta final states within the same kinematical range, i.e., $0<Q^2/{\rm GeV}^2<4$, $1.13<W/{\rm GeV}<1.6$. However, we emphasize that all two-body channels and all spin configurations of the three-body channels $\rho N$, $\pi\Delta$, and $p\sigma$ are considered in the intermediate states. The data coverage in this kinematical window is summarized in Fig.~\ref{fig:data}.

The new $\eta N$ data set consists entirely of  differential cross sections, given by
\begin{align}
    \frac{d\sigma}{d\Omega_{e}'dE_{\rm f}d\Omega}&=
    \left(
    \frac{\alpha}{2\pi^2}
    \frac{E_e'}{E_e}
    \frac{q_L}{Q^2}
    \frac{1}{1-\epsilon}\right)
    \frac{d\sigma^v}{d\Omega}\,,
\label{eq:diff-cross-sec.}
\end{align}
where $\Omega$ refers to the angles of the final meson-baryon system ($\theta$, $\phi$) and $\Omega_e'$ are the angles of the final electron at energy $E_e'$. The energy of the initial electron is denoted by $E_e$. The differential cross section $d\sigma^v/d\Omega$ for the virtual photon sub-process is commonly further decomposed as
\begin{align}
    \frac{d\sigma^v}{d\Omega}=&
    \sigma_{T}+\epsilon\sigma_{L}
    +\sqrt{2\epsilon(1+\epsilon)}\sigma_{LT}\cos{\phi}\nonumber\\
    &+\epsilon\sigma_{TT}\cos{2\phi}
    +h\sqrt{2\epsilon(1-\epsilon)}\sigma_{LT'}\sin{\phi}
\label{eq:DSG}\,.
\end{align}
In contrast, in both $\pi N$ channels also polarization data have been measured which are connected to the multipoles as described explicitly in Ref.~\cite{Mai:2021vsw}. As discussed there, when both differential cross section data and structure functions ($\sigma_T$, $\sigma_L$, $\sigma_{TT}$, $\sigma_{LT}$, and $\sigma_{LT'}$) were available from the same experiment at the same kinematics, double counting was avoided with preference given to the differential cross section data. This is statistically more sound, as correlations between separated contributions to the differential cross section data are typically not quoted.

\begin{table}[t]
\renewcommand{\arraystretch}{1.20}
\begin{tabular}{|p{.15\linewidth}|p{.3\linewidth}|p{.25\linewidth}|p{.24\linewidth}|}
    \hline
    Type&$N_{\rm data}^{\pi^0p}$&$N_{\rm data}^{\pi^+n}$&$N_{\rm data}^{\eta p}$\\
    \hline
    \hline
    $\rho_{LT}$
    &45~\cite{Mertz:1999hp,Elsner:2005cz}
    &--
    &--\\
    $\rho_{LT'}$
    &2644~\cite{Joo:2003uc, Sparveris:2002fh, Kelly:2005jj, Bartsch:2001ea,Bensafa:2006wr}
    &4354~\cite{Joo:2004mi, PARK-pc-08-2007}
    &-- \\
    $\sigma_L$
    &--
    &2~\cite{Gaskell:2001fn}
    &--\\
    $d\sigma/d\Omega$
    &39942~\cite{Laveissiere:2003jf, Ungaro:2006df, Gayler:1971zz, May:1971zza, Hill:1977sy, Joo:2001tw, Frolov:1998pw, Siddle:1971ug, Haidan:1979yqa, Sparveris:2002fh, Kelly:2005jj, Kalleicher:1997qf, Baetzner:1974xy, Latham:1979wea, Latham:1980my, Stave:2006jha, Sparveris:2006uk, Alder:1975xt, Afanasev:1975qa, Shuttleworth:1972nw, Blume:1982uh, Rosenberg:1979zm, Gerhardt:1979zz}
    &32813~\cite{Egiyan:2006ks, Breuker:1977vy, PARK-pc-08-2007, Bardin:1975oea, Bardin:1977zu, Gerhardt:1979zz, DavenportMartyn1980Eona, Vapenikova:1988fd, Hill:1977sy, Alder:1975na, Evangelides:1973gg, Breuker:1982nw, Breuker:1982um}
    &1874~\cite{Armstrong:1998wg, Denizli:2007tq, Thompson:2000by}\\
    $\sigma_T+\epsilon \sigma_L$
    &318~\cite{Laveissiere:2003jf, Mertz:1999hp, Sparveris:2002fh, Kunz:2003we, Stave:2006ea, Sparveris:2006uk, Sparveris:2004jn, Alder:1975xt}
    &144~\cite{Breuker:1977vy, Alder:1975na}
    &--\\
    $\sigma_{T}$
    &10~\cite{Blume:1982uh}
    &2~\cite{Gaskell:2001fn}
    &--\\
    $\sigma_{LT}$
    &312~\cite{Laveissiere:2003jf, Sparveris:2002fh, Mertz:1999hp, Kunz:2003we, Stave:2006ea, Sparveris:2006uk, Sparveris:2004jn, Alder:1975xt}
    &106~\cite{Breuker:1977vy, Alder:1975na}
    &--\\
    $\sigma_{LT'}$
    &198~\cite{Joo:2003uc, Kunz:2003we, Stave:2006ea, Sparveris:2006uk}
    &192~\cite{Joo:2003uc}
    &--\\
    $\sigma_{TT}$
    &266~\cite{Laveissiere:2003jf, Stave:2006ea, Sparveris:2006uk, Sparveris:2004jn, Alder:1975xt}
    &91~\cite{Breuker:1977vy, Alder:1975na}
    &--\\
    $K_{D1}$
    &1527~\cite{Kelly:2005jj}
    &--
    &--\\
    $P_Y$
    &--
    &2~\cite{Warren:1999pq, Pospischil:2000ad}
    &--\\
    \hline
\end{tabular}
\caption{Data used in the fit, separated by observable type and final state.}
\label{TAB:data_types}
\end{table}

To study constraints of the experimental data on the present coupled-channel formalism, we employ the following fit strategies. First, starting with the fit results of the pion-electroproduction analysis~\cite{Mai:2021vsw}, including S-, P- and D-waves, while setting $N=2$ in Eq.~\eqref{eq:formfactor-Ftilde}, we allow for 40 new parameters,
\begin{align}
&\beta^0_{\eta N}\,,~\beta^1_{\eta N}\,,~\beta^2_{\eta N}~,
&&\quad\text{for}\quad E_{\ell\pm},L_{\ell\pm},M_{\ell\pm}\text{~and~}\ell\leq 2 \,,\nonumber\\
&\zeta^{NP}_{\eta N}~,
&&\quad\text{for}\quad L_{1-}~,
\end{align}
in addition to the 209 previously~\cite{Mai:2021vsw} available parameters. Specifically, for all intermediate channels we chose again $\zeta^{NP}_{\mu\neq\eta N}\equiv \zeta^{NP}_{\pi N}$ and $ \beta_{\mu\notin\{\pi N,\eta N\}}^{i\in\{0,1,2\}}=0$. Second, starting from any of the four best fit results\footnote{These solutions were obtained following different fit strategies in order to obtain a representation of the systematic uncertainty. The two solutions of Ref.~\cite{Mai:2021vsw}, with extended $Q^2$ ranges of up to 8~GeV${}^2$, are not used in this work.} of Ref.~\cite{Mai:2021vsw} and holding all but the new parameters fixed, we minimize a regular $\chi^2$ function
\begin{align}
&\chi^2_{\rm reg}=\sum_{i=1}^{N_{\rm all}}\left(\frac{\mathcal{O}_i^{\rm exp}-\mathcal{O}_i}{\Delta_i^{\rm stat}+\Delta^{\rm syst}_i}\right)^2\,,
\label{eq:chi2tot}
\end{align}
with respect to the $\pi^0p,\pi^+n,\eta p$ channels simultaneously. The data, as taken from  SAID, contain also systematic uncertainties $\Delta^{\rm syst}$ that are separately quoted in the data base~\cite{SAID-web}. We note that the database sizes are vastly different in these channels. Thus, this simple choice of the $\chi^2$ function might marginalize the influence of the smaller $\eta N$ dataset. To test this hypothesis, we additionally perform a minimization with respect to a commonly used weighting scheme (for a typical application see, e.g., ~\cite{Mai:2012dt}),
\begin{align}
&\chi^2_{\rm wt}=
\sum_{j\in\{\pi^0p,\pi^+n,\eta p\}} \frac{N_{\rm all}}{3N_j}
\sum_{i=1}^{N_j}\left(\frac{\mathcal{O}_{ji}^{\rm exp}-\mathcal{O}_{ji}}{\Delta_{ji}^{\rm stat}+\Delta^{\rm syst}_{ji}}\right)^2\,.
\label{eq:chi2dem}
\end{align}
Third, after the minimization routine (utilizing MINUIT library~\cite{James:1994vla}) has converged, all $209+40=249$ parameters are relaxed and the minimization is repeated leading to the eight different solutions $\{\mathfrak{F}^{\rm reg}_1,...,\mathfrak{F}^{\rm reg}_4,\mathfrak{F}^{\rm wt}_1,...,\mathfrak{F}^{\rm wt}_4\}$, discussed in the next section.

\section{Results}
\label{sec:results}

\begin{table}[t]
\renewcommand{\arraystretch}{1.50}
\begin{tabular}{|p{.20\linewidth}|p{.25\linewidth}|p{.15\linewidth}|p{.15\linewidth}|p{.15\linewidth}|}
\hline
&$\chi^2/{\rm dof}$&$\chi^2_{\pi^0p/\rm data }$&$\chi^2_{\pi^+n/\rm data}$&$\chi^2_{\eta p/\rm data}$\\
\hline\hline
$\mathfrak{F}^{\rm reg}_1$ &1.66& 1.68& 1.61& 1.77\\
$\mathfrak{F}^{\rm reg}_1$ &1.73& 1.71& 1.71& 2.29\\
$\mathfrak{F}^{\rm reg}_1$ &1.69& 1.69& 1.66& 1.89\\
$\mathfrak{F}^{\rm reg}_1$ &1.69& 1.7& 1.64& 2.05\\
\hline
$\mathfrak{F}^{\rm wt}_1$ &1.54& 1.74& 1.63& 1.25\\
$\mathfrak{F}^{\rm wt}_1$ &1.63& 1.82& 1.79& 1.27\\
$\mathfrak{F}^{\rm wt}_1$ &1.58& 1.74& 1.73& 1.27\\
$\mathfrak{F}^{\rm wt}_1$ &1.58& 1.79& 1.6& 1.33\\
\hline
\end{tabular}
\caption{Fit results of the coupled-channel JBW analysis with respect to $\pi N$ and $\eta N$ data. The second column shows results of a fit using standard~\eqref{eq:chi2tot} ('reg') and weighted~\eqref{eq:chi2dem} ('wt')  $\chi^2$ functions, respectively, whereas the last three columns separate out contributions for individual final-state channels (per datum).}
\label{tab:fit-results-x}
\end{table}


Each of the followed fit strategies led to a successful description of both considered channels. The fit results are collected in  Table~\ref{tab:fit-results-x} including contributions separated out for each of the considered final-state channels ($\pi^0p$, $\pi^+n$, $\eta p$). As expected, fit results relying on the weighted version of the $\chi^2$ function~\eqref{eq:chi2dem} led to a much better description of the $\eta p$ data, which are much sparser than the $\pi N$ data. When comparing the individual contributions to those of the previous JBW single-channel study~\cite{Mai:2021vsw} we note that the description of both $\pi N$ channels is similar in the present coupled-channel analysis. The same holds true for the contributions to subsets of data separated with respect to individual observable types. For more details on the $\pi N$ channels, see Ref.~\cite{Mai:2021vsw} as well as the interactive JBW homepage~\cite{JBW-homepage}.

\begin{figure}[t]
    \includegraphics[height=4.1cm,trim=0 1.2cm 0 0,clip]{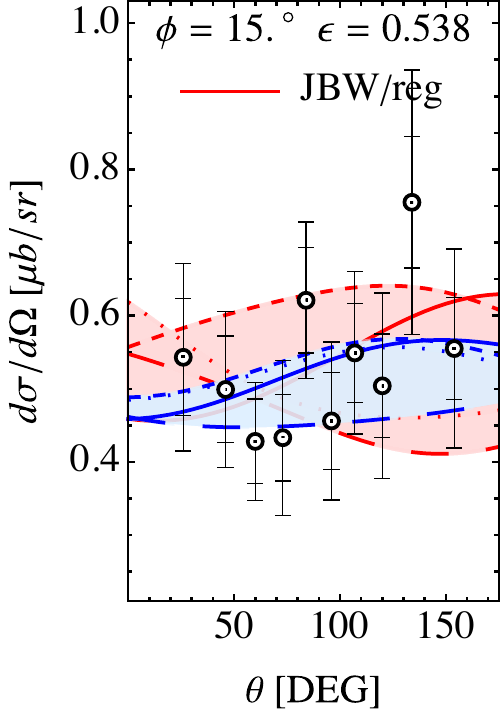}
    \includegraphics[height=4.1cm,trim=1.3cm 1.2cm 0 0,clip]{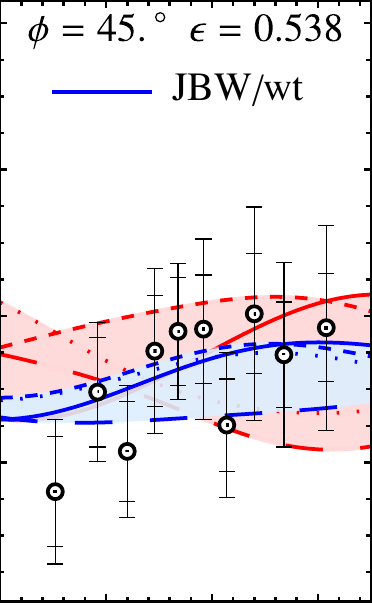}
    \includegraphics[height=4.1cm,trim=1.3cm 1.2cm 0 0,clip]{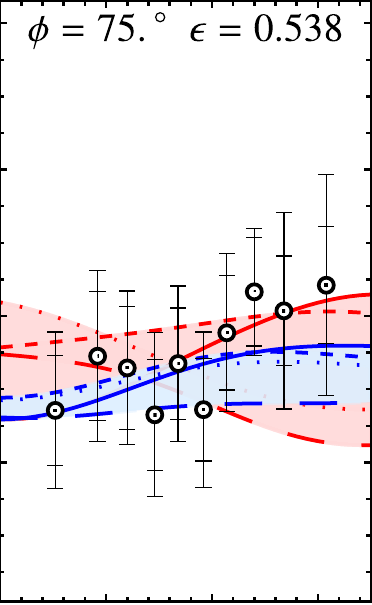}\\
    \includegraphics[height=4.9cm,trim=0 0 0 0,clip]{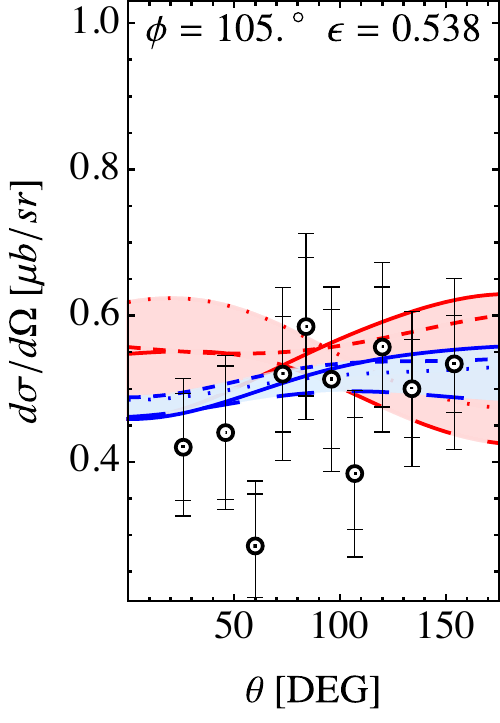}
    \includegraphics[height=4.9cm,trim=1.3cm 0 0 0,clip]{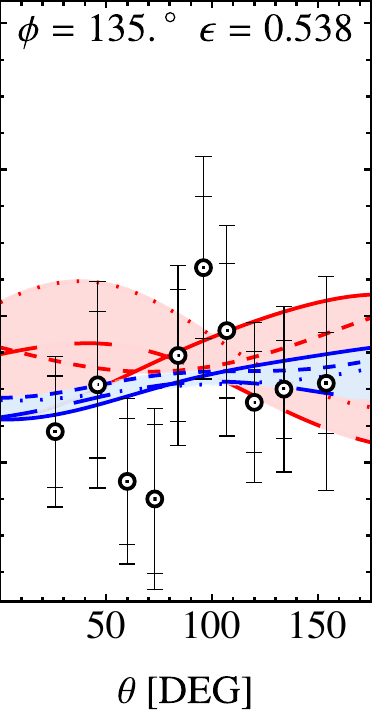}
    \includegraphics[height=4.9cm,trim=1.3cm 0 0 0,clip]{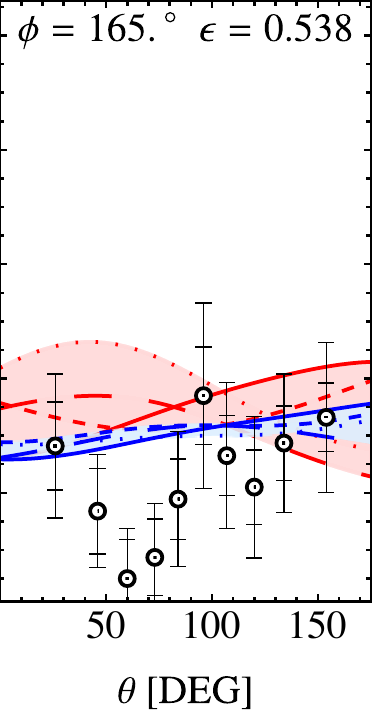}
    \caption{Selected subset of $\eta N$ data for $W=1.5~{\rm GeV}$, $Q^2=1.2~{\rm GeV}^2$ from Ref.~\cite{Denizli:2007tq}. The four red and blue lines, respectively, correspond to the ``reg'' and ``wt'' solutions of Table~\ref{tab:fit-results-x}.
    \label{fig:eta-DSG-1}}
\end{figure}

\begin{figure}[t]
    \includegraphics[height=4.7cm,trim=0 0 0 0,clip]{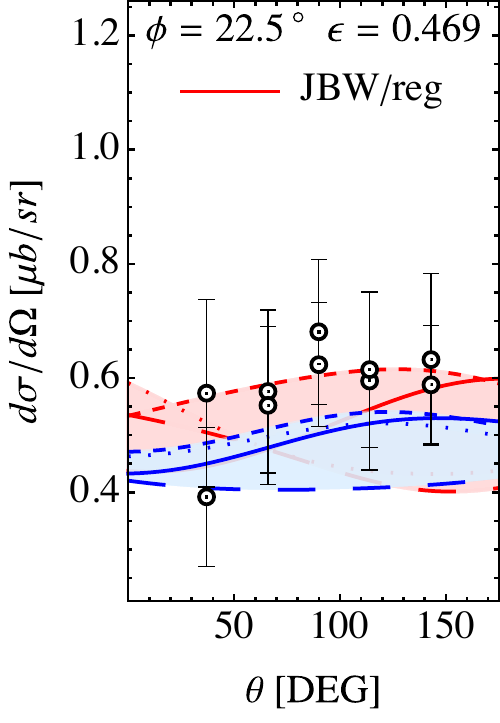}
    \includegraphics[height=4.7cm,trim=1.2cm 0 0 0,clip]{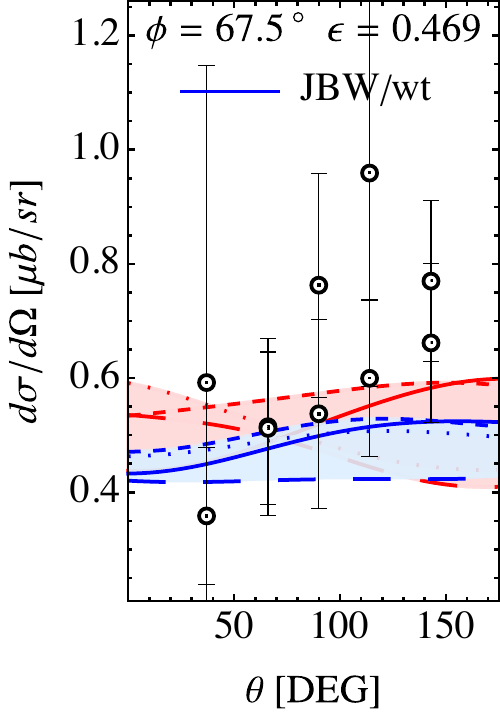}
    \includegraphics[height=4.7cm,trim=1.2cm 0 0 0,clip]{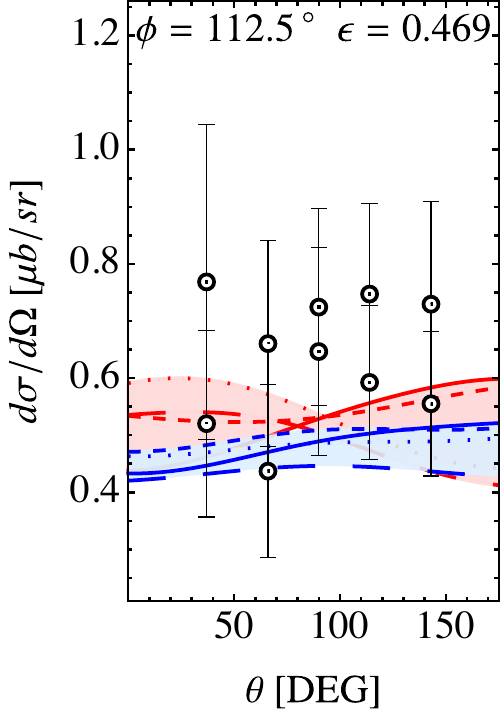}
    \caption{Selected subset of $\eta N$ data for $W=1.5~{\rm GeV}$, $Q^2=1.25~{\rm GeV}^2$ from Ref.~\cite{Thompson:2000by, RAThompson2000}.
    The four red and blue lines, respectively, correspond to the ``reg'' and ``wt'' solutions of Table~\ref{tab:fit-results-x}.
    \label{fig:eta-DSG-2}}
\end{figure}

\begin{figure}[t]
    \includegraphics[height=3.82cm,trim=0 1.5cm 0 0,clip]{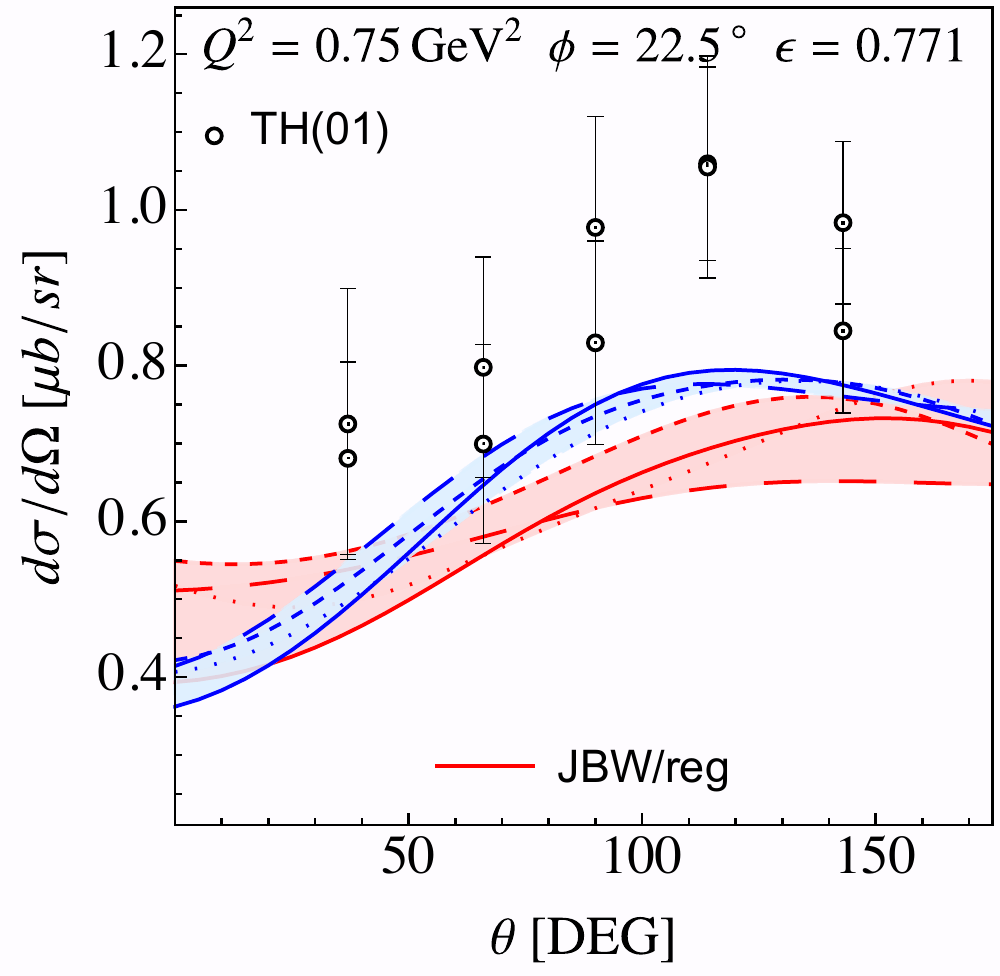}
    \includegraphics[height=3.82cm,trim=1.75cm 1.5cm 0 0,clip]{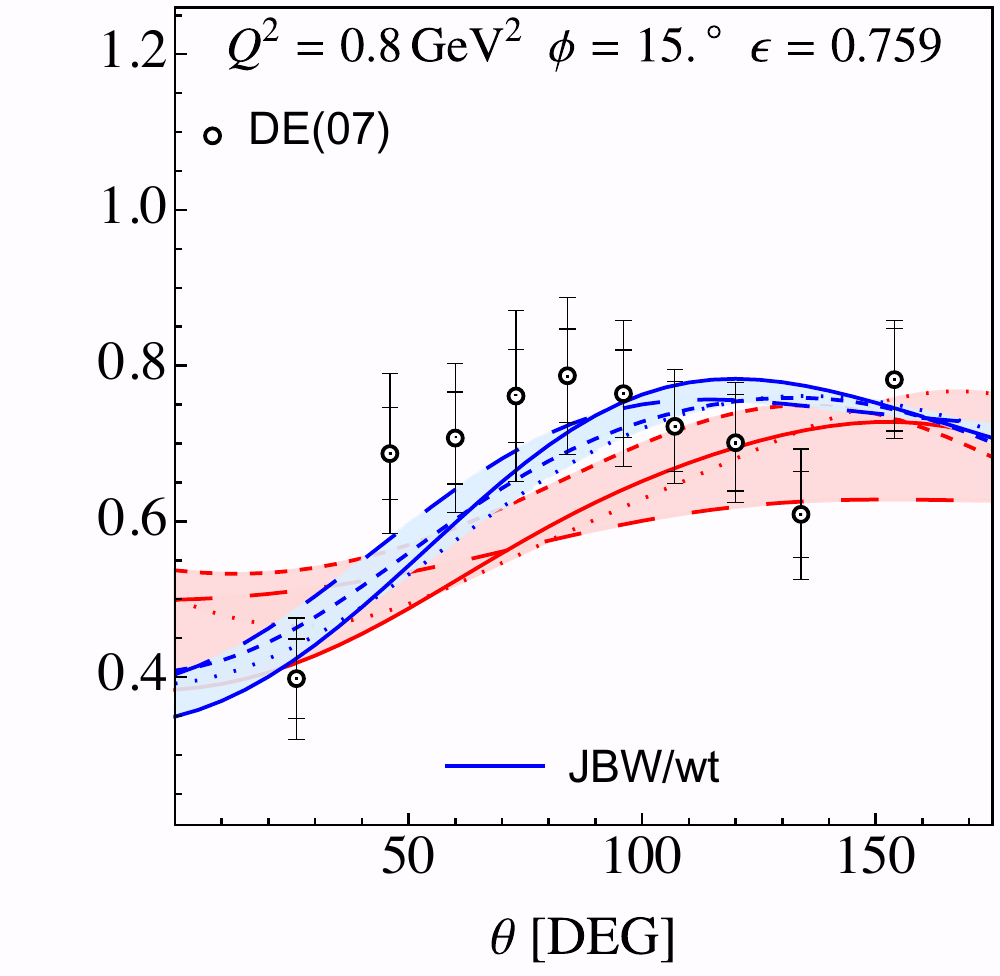}\\
    \includegraphics[height=4.5cm,trim=0 0 0 0,clip]{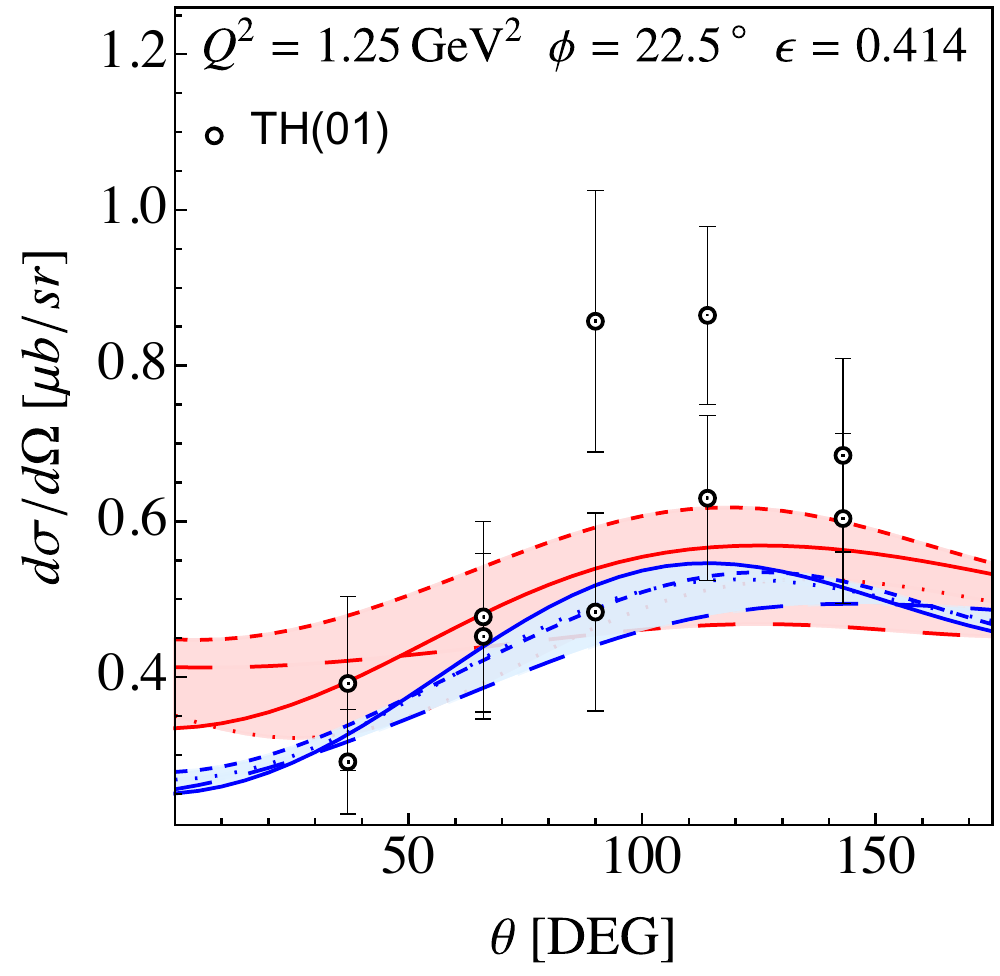}
    \includegraphics[height=4.5cm,trim=1.75cm 0 0 0,clip]{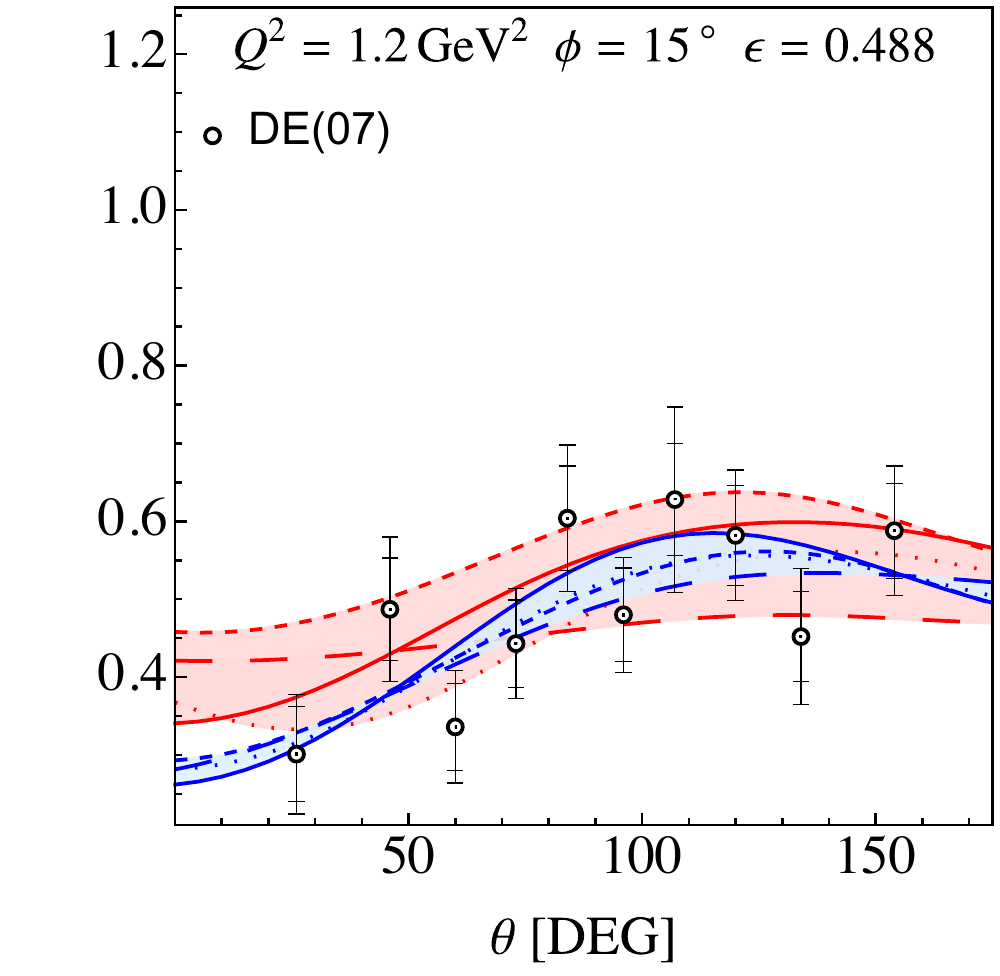}
    \caption{Selected subset of $\eta N$ data for $W=1.56~{\rm GeV}$ from Refs. TH(01)~\cite{Thompson:2000by, RAThompson2000} and DE(07)~\cite{Denizli:2007tq} for similar kinematics. Result of the JBW coupled-channel fits are shown by red and blue lines corresponding to the eight best fit solutions.
    \label{fig:eta-DSG-3}}
\end{figure}

Taking a closer look on the fit results we find a relatively weak $\phi$-dependence in the data and corresponding fits, see Figs.~\ref{fig:eta-DSG-1} and \ref{fig:eta-DSG-2}. In the latter figure, there are two data points from Ref.~\cite{RAThompson2000} for each angle $\theta$ at a fixed
value of $\phi$. These were obtained from measurements at azimuthal angles $\phi$ and $(360-\phi)$, respectively. The cross sections were averaged in Ref.~\cite{Thompson:2000by}, but both values are retained in our database. In Fig.~\ref{fig:eta-DSG-3} we compare fits and data at nearby kinematic points for which the fit curves are nearly identical. This gives a visual comparison of the data consistency.

\begin{figure*}[t]
    \includegraphics[width=0.98\linewidth,trim=0.cm 0.7cm 0.8cm 0,clip]{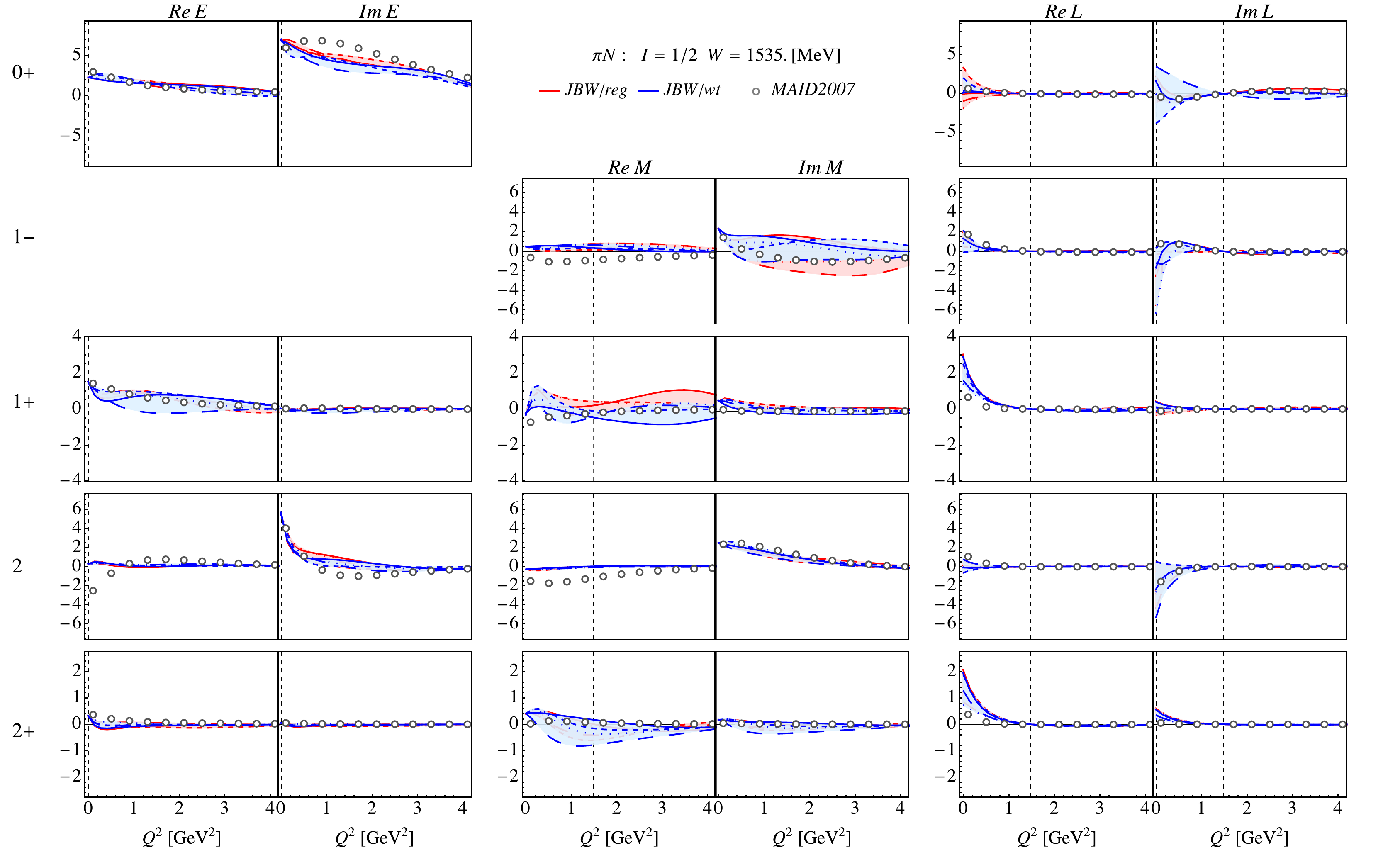}\\
    \includegraphics[width=0.98\linewidth,trim=0.cm 0 0.8cm 0cm,clip]{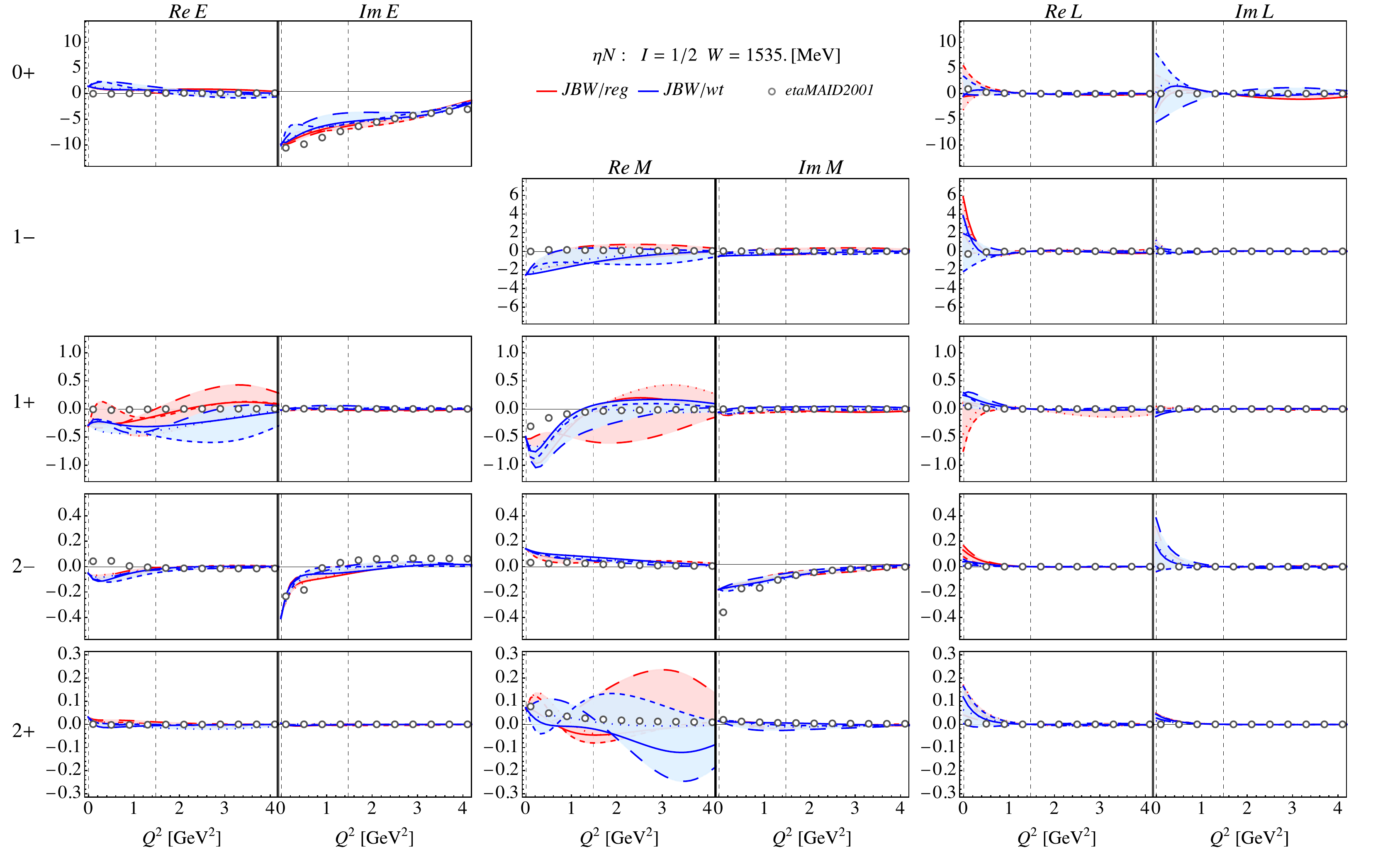}
    \caption{JBW coupled-channel results. Multipoles (in mfm) in the $\pi N$ and $\eta N$ $I=1/2$ channel at $W=1535~{\rm MeV}$ in comparison to the result of the MAID2007~\cite{Drechsel:2007if} and etaMAID2001~\cite{Chiang:2001as} analyses, respectively. The latter results are extracted from the MAID homepage and multiplied by $-1/\sqrt{3}$, adjusting for a phase- and isospin factor. The leftmost column shows the total angular momentum. Fits correspond to the results of Tab.~\ref{tab:fit-results-x} with the same line shape coding as in Ref~\cite{Mai:2021vsw}.
    \label{fig:mult-1535}
    }
\end{figure*}

As for underlying multipoles, we found that in most cases longitudinal multipoles are subdominant to electric and magnetic ones. An overview of all considered multipoles is shown in Fig.~\ref{fig:mult-1535} for the c.m. energy fixed to 1535~MeV. There, in most cases and within the systematic uncertainties of our approach --- quantified by the spread of predictions from fits  $\{\mathfrak{F}^{\rm reg}_1,...,\mathfrak{F}^{\rm reg}_4,\mathfrak{F}^{\rm wt}_1,...,\mathfrak{F}^{\rm wt}_4\}$ --- we observe an agreement with the MAID2007 ($\pi N$) and etaMAID ($\eta N$) multipole predictions\footnote{A more quantitative statement is impossible due to missing uncertainty estimations for the etaMAID parametrizations}. The isospin $I=1/2$ $\pi N$ multipoles are shown in the same figure with the pertinent comparison to the MAID2007 solution for convenience.

Fixing the virtuality $Q^2$ to some values of interest the multipoles are shown in Figs.~\ref{fig:eta-mult-Q200/1000} and~\ref{fig:eta-mult-Q2000/3000}. There, we observe that the dominant $E_{0+}$ multipole agrees well with that of the etaMAID parametrization when correcting for the phase convention ($-1$) and isospin factor ($1/\sqrt{3}$). To be clear, we show our multipoles in the isospin basis, which make them smaller by a factor of $1/\sqrt{3}$ compared to the etaMAID multipoles which are quoted in the particle basis. As the results show, longitudinal multipoles seem indeed very small compared to the electric and magnetic ones. Interestingly, the $M_{2-}$ multipole seems to have a similar trend as that of the MAID solution, while the corresponding uncertainties seem to change with different $Q^2$ values. This can be attributed to the gaps in $\eta p$ data at some fixed $Q^2$ kinematics, see next section.

\begin{figure*}[t]
    \includegraphics[width=0.45\linewidth,trim=0 0cm 0 0cm,clip]{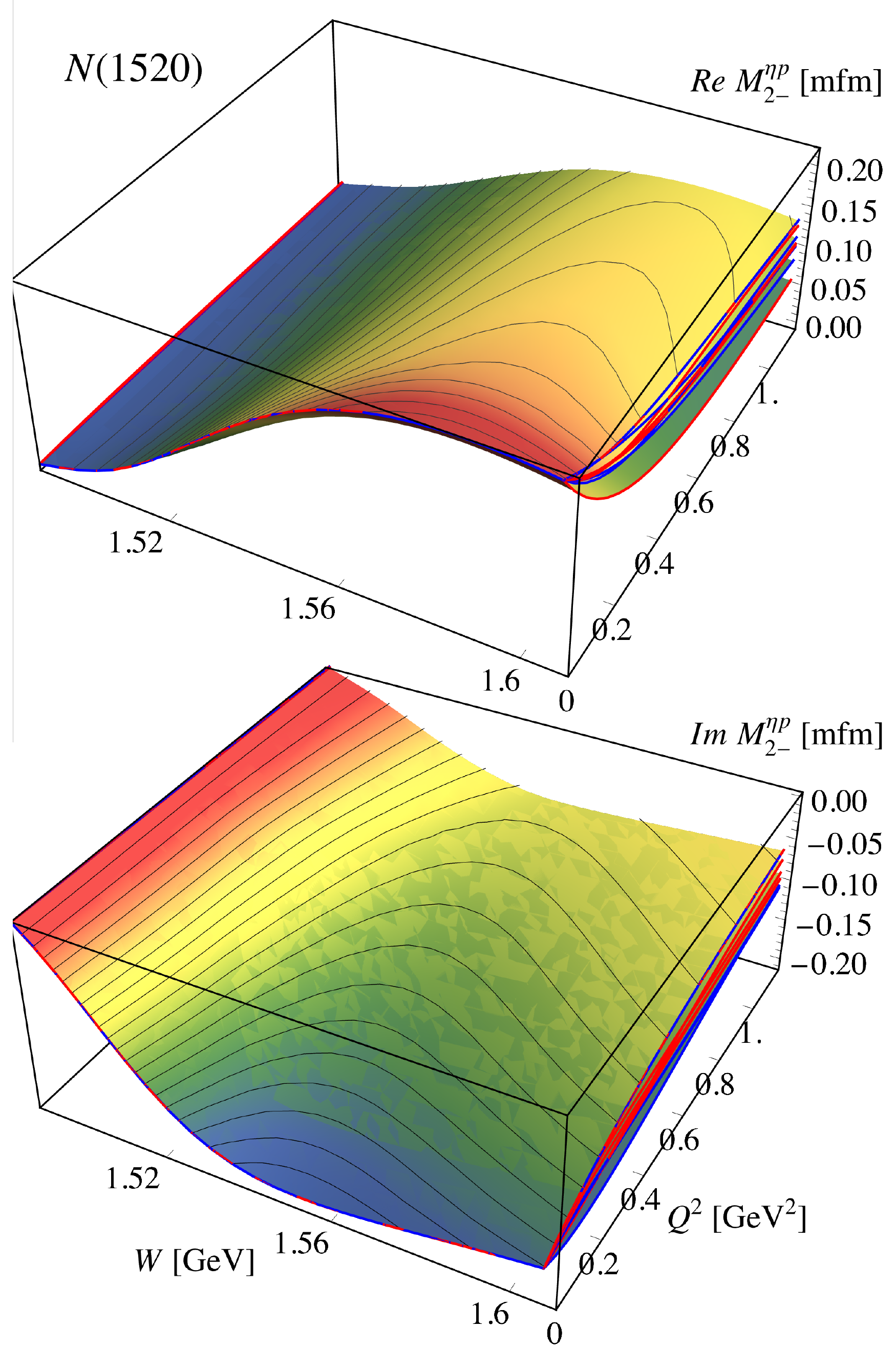}
    \includegraphics[width=0.45\linewidth,trim=0 0cm 0 0cm,clip]{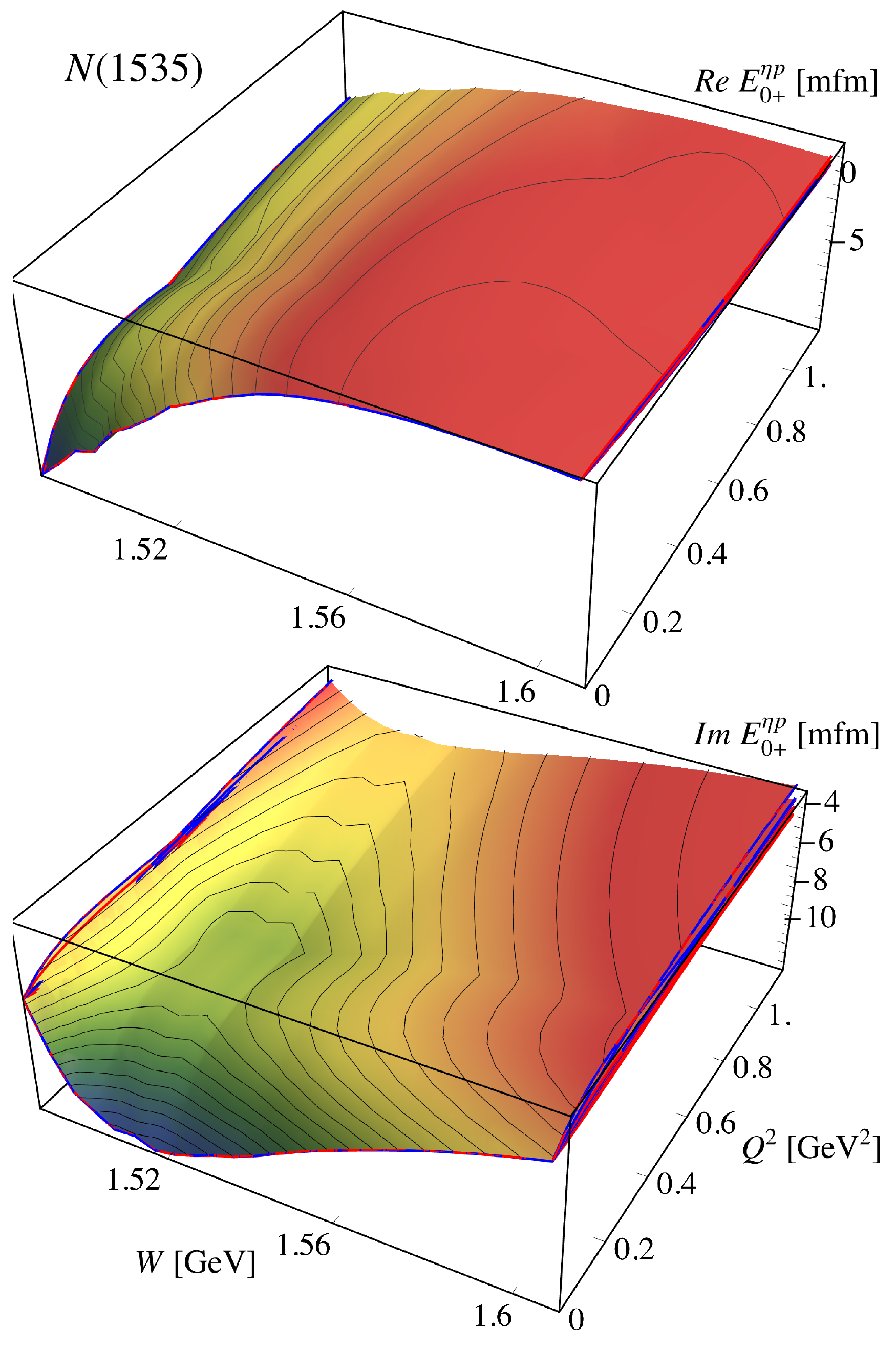}
    \caption{Predictions of $E_{0+}(W,Q^2)$ corresponding to the quantum numbers of $N(1535)$. Different surfaces correspond to the best fit solutions obtained minimizing the regular (red boundary lines) and weighted (blue boundary lines) version of the $\chi^2$ function.
    \label{fig:3d-1535}
    }
\end{figure*}

Finally, we demonstrate in Fig.~\ref{fig:3d-1535} the full $Q^2$ vs $W$ dependence of the $E_{0+}$ and $M_{2-}$ multipoles, corresponding to quantum numbers of the $N(1535)1/2^-$ and $N(1520)3/2^-$. We observe that the systematic uncertainties discussed above are well under control. In particular, all fit solutions show a non-trivial $Q^2$ dependence. This supports our expectation that the helicity couplings will carry new physical information, when full $(W,Q^2)$ information is extracted, being part of our future plans.

\section{Discussion and Conclusion}
\label{sec:conclusion}
We have generalized our recent analysis of pion electroproduction~\cite{Mai:2021vsw} to include eta electroproduction data. This allowed a coupled-channel fit up to $W=$1.6~GeV for $Q^2<4$ GeV$^2$. For both reactions, partial waves up to $\ell=2$ were included. Given that the pion and eta electroproduction databases are very different in size, we compared $\chi^2$ minimization without and with weighting factors to increase the influence of the smaller eta electroproduction dataset.

As in Ref.~\cite{Mai:2012wy}, using different fit strategies, we found several solutions with nearly equivalent $\chi^2$/data values. The fits achieved $\chi^2$/data values near 1.7, similar to our previous fits to pion data alone. The fit to pion data and the resulting multipoles showed little change from the single-channel case. This result held in both weighted and unweighted fits. 

Also, as in our pion electroproduction fits, the spread of results for multipoles provided a measure of systematic errors. As expected, the $E_{0+}$ multipole was reliably determined with a $Q^2$ dependence similar to that exhibited in  etaMAID (once a phase ambiguity was accounted for).

The evidence for contributions from higher partial waves depends on the experiment. As mentioned, the data of Ref.~\cite{Armstrong:1998wg} are compatible with a Breit-Wigner contribution from the $N(1535)$, without any need for a background term, higher partial waves, or longitudinal multipoles, for a fit covering the energy range considered in the present analysis. The later experiment of Ref.~\cite{Denizli:2007tq}, however, displays a clear forward-backward asymmetry (a sign of P-wave interference) and some evidence for D-wave interference producing a convex shape. 

In our multipole solutions, there is evidence for sizeable P-wave contributions, but the spread implies the P-wave multipoles are not well determined. This is expected since the corresponding candidate states $N(1710)1/2^+$ and $N(1720)3/2^+$ are beyond the upper energy limit. Interestingly, the $M_{2-}$ multipoles are quite consistent, and appear to give a consistent value for this multipole, which also agrees with the etaMAID values away from the $Q^2=0$ photon point and the upper $Q^2$ limit of our fits. 

\begin{figure*}[t]
    \includegraphics[width=\linewidth,trim=0cm 0.7cm 0.8cm 0,clip]{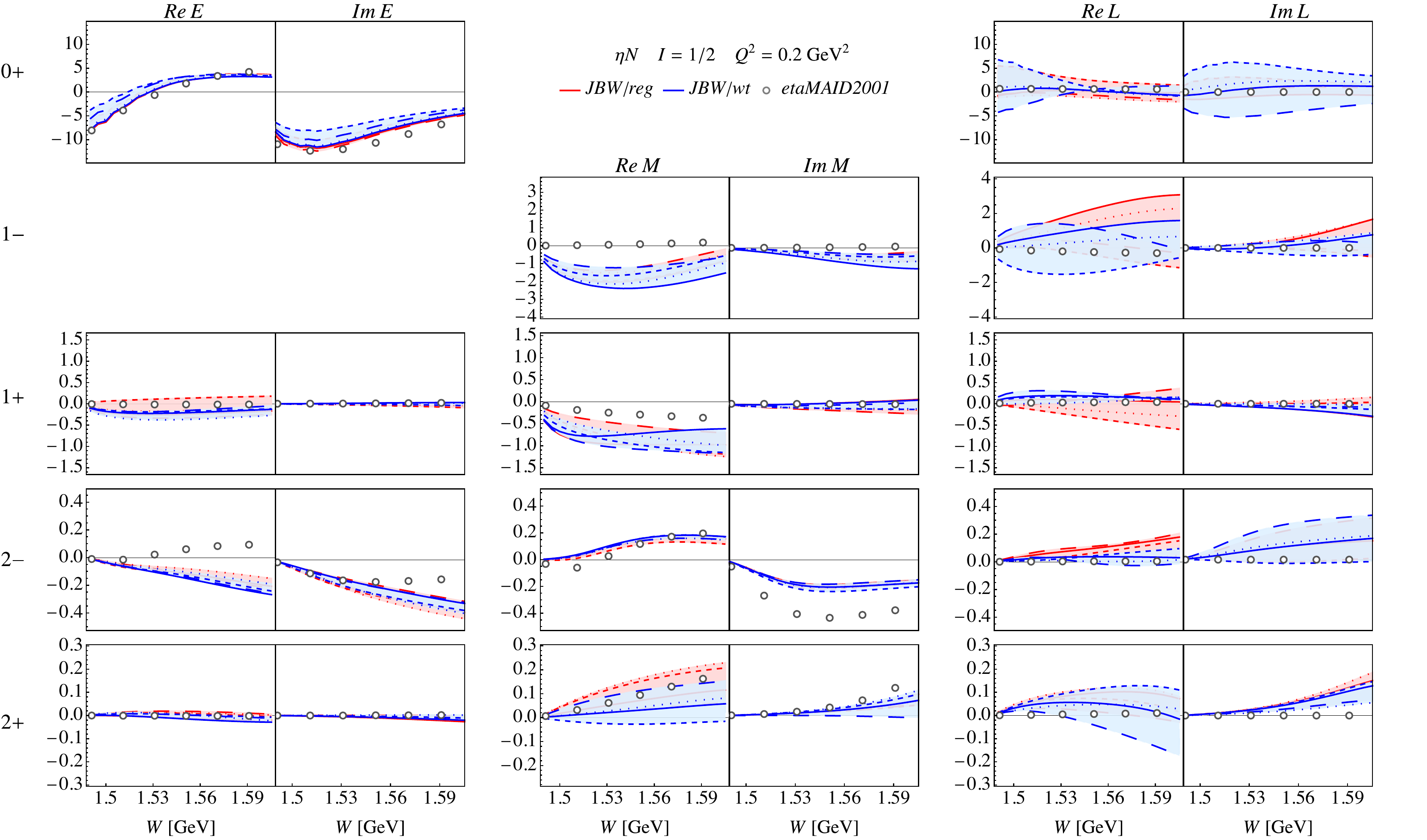}
    \includegraphics[width=\linewidth,trim=0cm 0 0.8cm 0.45cm,clip]{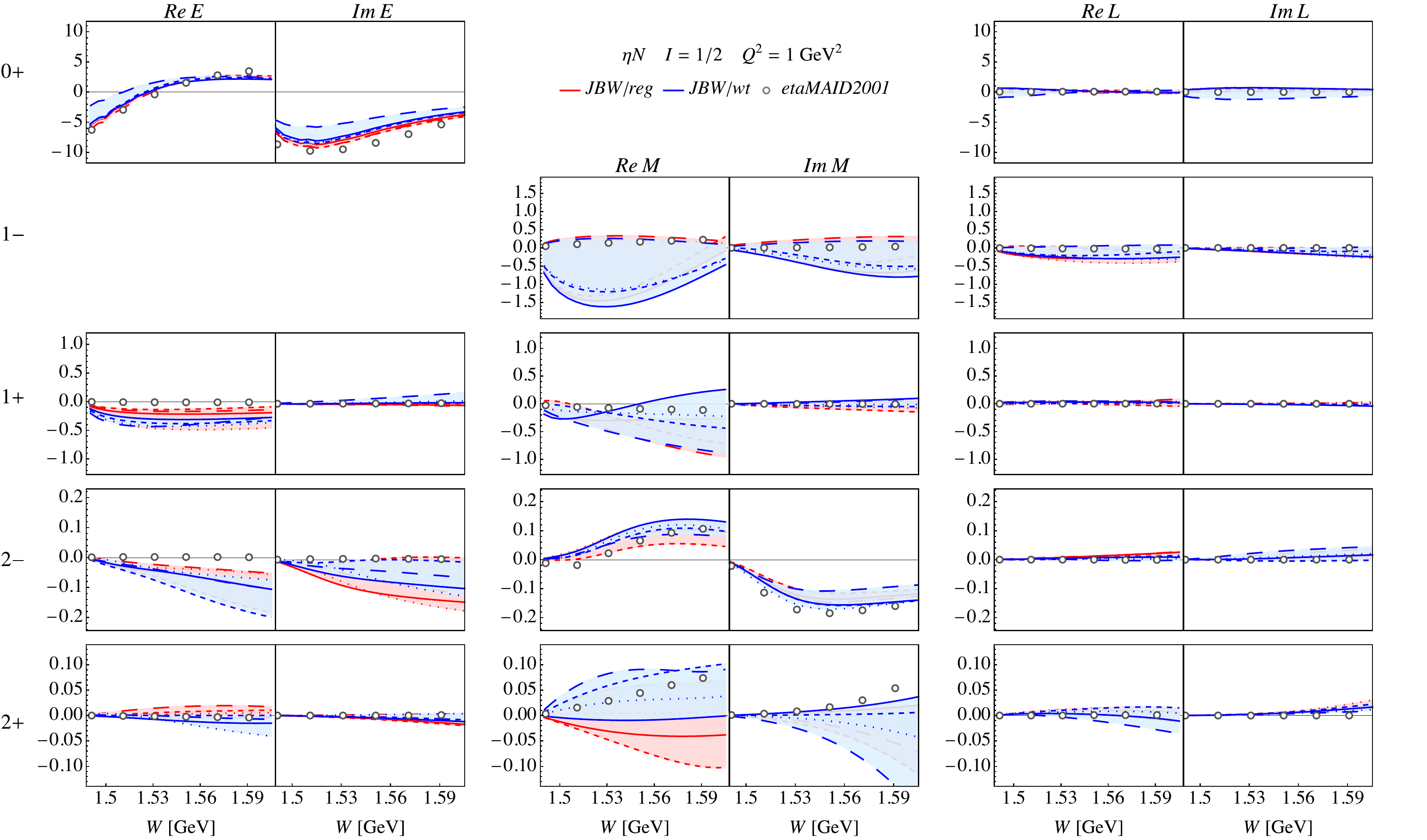}
    \caption{Eta electroproduction multipoles for lower fixed values of $Q^2$. Notation as given in Figure~\ref{fig:mult-1535}.
    \label{fig:eta-mult-Q200/1000}
    }
\end{figure*}

\begin{figure*}[t]
    \includegraphics[width=\linewidth,trim=0cm 0.7cm 0.8cm 0,clip]{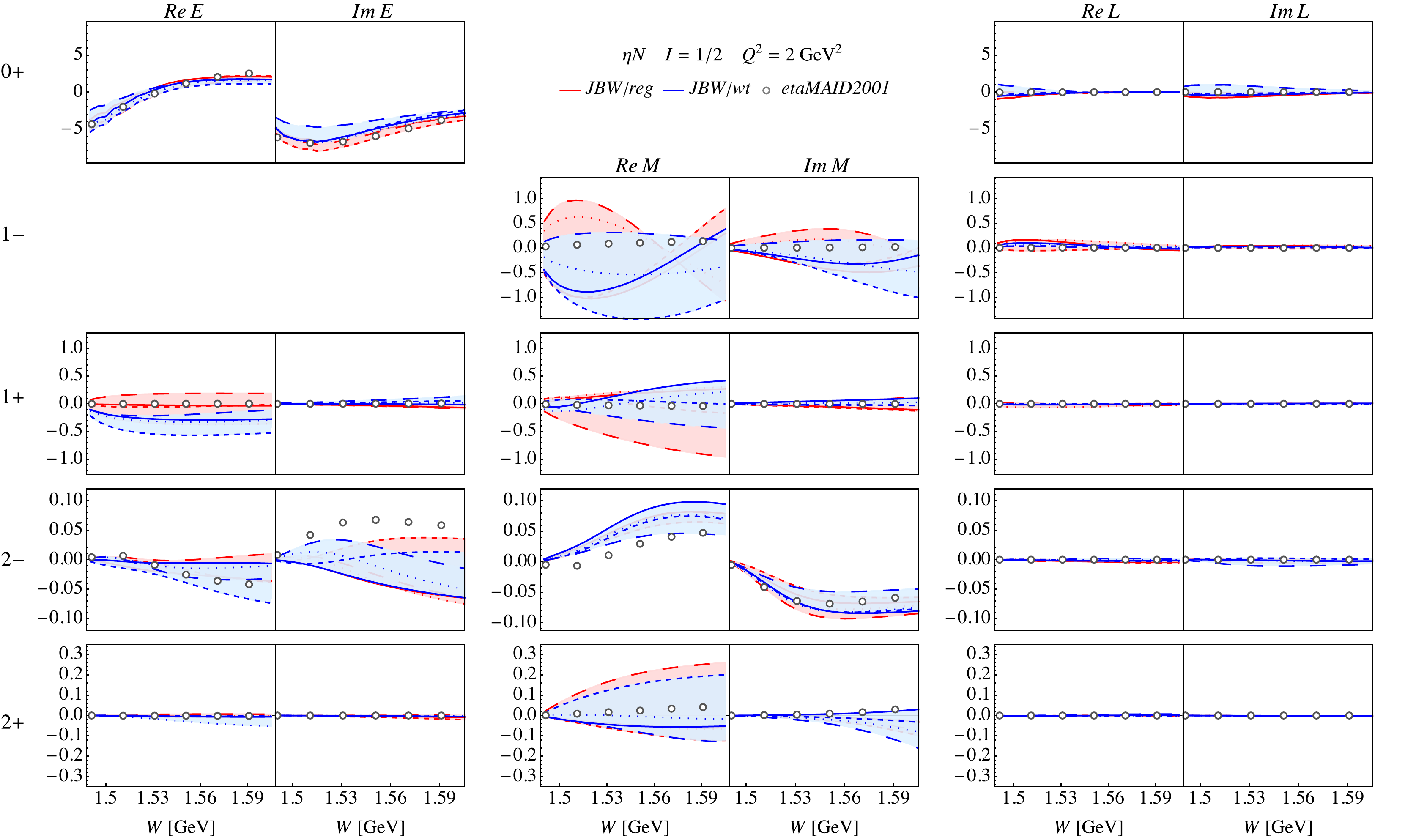}\\[3mm]
    \includegraphics[width=\linewidth,trim=0cm 0 0.8cm 0.45cm,clip]{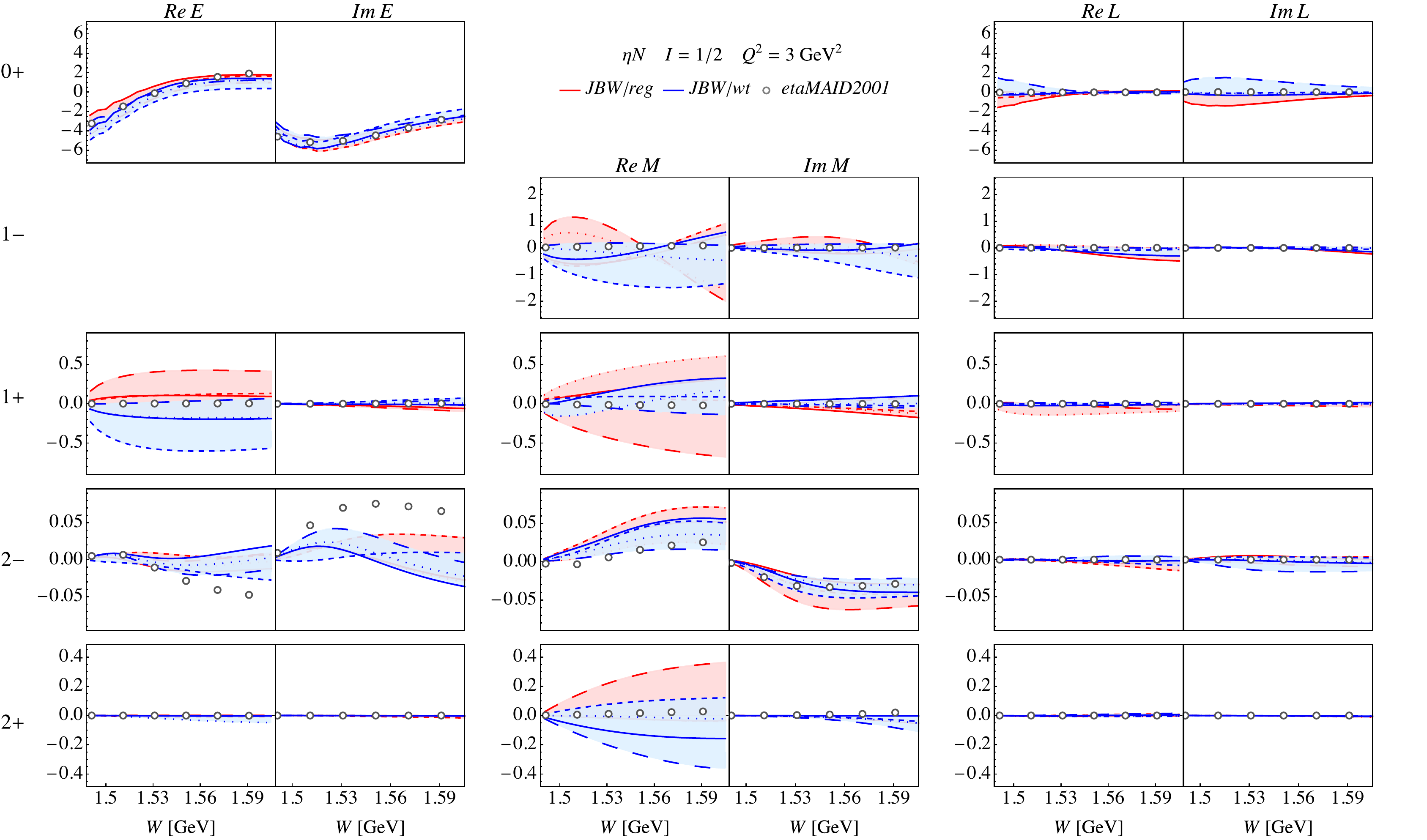}
    \caption{Eta electroproduction multipoles for higher fixed values of $Q^2$. Notation as given in Figure~\ref{fig:mult-1535}.
    \label{fig:eta-mult-Q2000/3000}}
\end{figure*}

We can understand the consistency of multipole determinations in Figs.~\ref{fig:eta-mult-Q200/1000} and~\ref{fig:eta-mult-Q2000/3000}, plotting multipoles versus center-of-mass energy for fixed  $Q^2$ (0.2, 1, 2, and 3 GeV$^2$), based on the data fitted and the constraint at $Q^2=0$. The lowest-$Q^2$ plot displays multipoles for a $Q^2$ value below the lower limit (0.3 GeV$^2$) of fitted data and is close to the photoproduction point, where it was shown that the etaMAID and J\"uBo fits can be quite different. The behavior at $Q^2=3$\,GeV$^2$ is supported by data from Ref.~\cite{Armstrong:1998wg} (2.4 and 3.6 GeV$^2$) which can be fitted with only a Breit-Wigner contribution to $E_{0+}$ and shows no evidence for P- and D-waves.

The plots for intermediate $Q^2$ are supported by data from Refs.~\cite{Thompson:2000by, Denizli:2007tq}, of which Ref.~\cite{Denizli:2007tq} is more precise. These plots show the most consistency for multipoles that can be determined over this narrow energy range. Note also that longitudinal multipoles are small (consistently among all our solutions) whereas this feature was built in in some earlier fits~\cite{Armstrong:1998wg}.

In summary, this first coupled-channel fit to both pion and eta electroproduction data supports an expanded study. As a first step, the number of included partial-waves and the energy limits will be increased.
Once completed, we will attempt an expansion to kaon electroproduction~\cite{CLAS:2002zlc, CLAS:2006ogr, Nasseripour:2008aa, CLAS:2009sbn, Carman:2012qj, CLAS:2014udv, Achenbach:2017pse}, in the near-threshold region. We will also be in a position to explore resonance behavior at the pole as a function of $Q^2$ using tools developed to study the J\"uBo photoproduction amplitudes.



\begin{center}
{\bf Acknowledgements}
\end{center}
This work is supported by the U.S. Department of Energy, DOE Office of Science, Office of Nuclear Physics awards DE-SC0016582 and DE-SC0016583 and  contract DE-AC05-06OR23177. It is also supported by the NSFC and the Deutsche Forschungsgemeinschaft (DFG, German Research Foundation) through the funds provided to the Sino-German Collaborative Research Center TRR110 “Symmetries and the Emergence of Structure in QCD” (NSFC Grant No.\ 12070131001, DFG Project-ID 196253076-TRR 110), by the Chinese Academy of Sciences (CAS) through a President's International Fellowship Initiative (PIFI) (Grant No. 2018DM0034), by the VolkswagenStiftung (Grant No.\ 93562), and by the EU Horizon 2020 research and innovation programme, STRONG-2020 project
under grant agreement No. 824093. The multipole calculation and parameter optimization are performed on the Colonial One computer cluster~\cite{Colonial-One}. The authors gratefully acknowledge the computing time granted through JARA on the supercomputer JURECA~\cite{jureca} at Forschungszentrum Jülich that was used to produce the input at the photon point.

\bibliography{BIB3,NON-INSPIRE}

\end{document}